\documentclass[reprint, 10pt, amsmath, amssymb, floatfix, aps, prd,  superscriptaddress, nofootinbib, nobibnotes,twocolumn]{revtex4-1}

\pdfoutput=1
\usepackage{graphicx}
\graphicspath{{figures/}}
\usepackage[utf8]{inputenc}
\usepackage{amsmath}
\usepackage{amsfonts}
\usepackage{amssymb}
\usepackage{multirow}
\usepackage{tcolorbox}
\usepackage{CJKutf8}
\usepackage{color}
\usepackage[colorlinks=true, linkcolor=red, citecolor=blue]{hyperref}
\usepackage{orcidlink}
\usepackage[capitalise]{cleveref}
\usepackage{acro}
\usepackage{adjustbox}
\usepackage{array}
\usepackage{natbib}
\usepackage{dblfnote}
\DFNalwaysdouble
\usepackage{slashed}
\usepackage{dcolumn}
\usepackage{hyperref}
\usepackage{geometry}
\usepackage{subfig} 
\usepackage{overpic}
\usepackage{placeins}
\usepackage{inputenc}
\usepackage{caption}
\newcommand*{\dif}{\mathop{}\!\mathrm{d}}

\def\({\left(}
\def\){\right)}
\def\[{\left[}
\def\]{\right]}
\def\be{\begin{eqnarray}}
\def\ee{\end{eqnarray}}

\DeclareAcronym{GW}{
  short = GW ,
  long = gravitational wave ,
  short-plural = s 
}
\DeclareAcronym{LIGO}{
  short = LIGO ,
  long = Laser Interferometer Gravitational-wave Observatory ,
  short-plural = 
}
\DeclareAcronym{LISA}{
  short = LISA ,
  long = Laser Interferometer Space Antenna ,
  short-plural =  
}
\DeclareAcronym{SKA}{
  short = SKA ,
  long = Square Kilometre Array ,
  short-plural =  
}  
  
\DeclareAcronym{SNR}{
	short = SNR ,
	long = signal-to-noise ratio ,
	short-plural = 
}

\DeclareAcronym{PTA}{
	short = PTA ,
	long = pulsar timing array ,
	short-plural = 
}

\DeclareAcronym{FLRW}{
  short = FLRW ,
  long = Friedmann-Lemaitre-Robertson-Walker ,
  short-plural =  
}

\DeclareAcronym{SIGW}{
	short = SIGW ,
	long = scalar induced gravitational wave ,
	short-plural =  s
}

\DeclareAcronym{PBH}{
	short = PBH ,
	long = primordial black hole ,
	short-plural =  s
}

\DeclareAcronym{SMBHB}{
  short = SMBHB ,
  long = supermassive black hole binary ,
  short-plural = s
}

\DeclareAcronym{KDE}{
  short = KDE ,
  long = kernel density estimator ,
  short-plural = s
}

\DeclareAcronym{BPBHM}{
  short = BPBHM ,
  long = binary primordial black hole merger,
  short-plural = s
}

\DeclareAcronym{CMB}{
	short = CMB ,
	long = cosmic microwave background ,
	short-plural =  
}
\DeclareAcronym{DM}{
	short = DM ,
	long = dark matter ,
	short-plural =  
}

\DeclareAcronym{BBN}{
	short = BBN ,
	long = Big-Bang nucleosynthesis ,
	short-plural =  
}

\DeclareAcronym{LN}{
	short = LN ,
	long = log-normal  ,
	short-plural =  
}

\DeclareAcronym{BPL}{
	short = BPL ,
	long = broken power-law ,
	short-plural =  
}

\DeclareAcronym{SGWB}{
	short = SGWB ,
	long = stochastic gravitational	wave background ,
	short-plural =  s
}

\DeclareAcronym{LSS}{
	short = LSS ,
	long = large scale structure ,
	short-plural =  
}

\DeclareAcronym{RD}{
	short = RD ,
	long = radiation-dominated ,
	short-plural =  
}

\DeclareAcronym{PLS}{
	short = PLS ,
	long = power low sensitivity ,
	short-plural =  
}

\DeclareAcronym{MAP}{
	short = MAP ,
	long = maximum a posterior ,
	short-plural =  
}

\DeclareAcronym{BAO}{
	short = BAO ,
	long = baryon acoustic oscillations ,
	short-plural = 
}

\begin{document}
%\begin{CJK}{UTF8}{}

\title{Cosmological constraints on small-scale primordial non-Gaussianity }

\author{Jing-Zhi Zhou\orcidlink{0000-0003-2792-3182}} 
\email{zhoujingzhi@tju.edu.cn}
\affiliation{Center for Joint Quantum Studies and Department of Physics,
School of Science, Tianjin University, Tianjin 300350, China}

\author{Zhi-Chao Li\orcidlink{0009-0005-7984-2626}}
\email{lizc@tju.edu.cn}
\affiliation{Center for Joint Quantum Studies and Department of Physics,
School of Science, Tianjin University, Tianjin 300350, China}

\author{Di Wu\orcidlink{0000-0001-7309-574X}}
\email{wudi@ucas.ac.cn}
\affiliation{School of Fundamental Physics and Mathematical Sciences, Hangzhou Institute for Advanced Study, University of Chinese Academy of Sciences, Hangzhou 310024, China}

\begin{abstract}
In contrast to the large-scale primordial power spectrum $\mathcal{P}_{\zeta}(k)$ and primordial non-Gaussianity $f_{\mathrm{NL}}$, which are strictly constrained, the small-scale $\mathcal{P}_{\zeta}(k)$ and $f_{\mathrm{NL}}$ are not as limited. Considering local-type primordial non-Gaussianity, we study the \acp{PBH} and \acp{SIGW} caused by large-amplitude small-scale primordial power spectrum. By analyzing current observational data from \ac{PTA}, \ac{CMB}, \ac{BAO}, and abundance of \acp{PBH}, and combining it with the \ac{SNR} analysis of \ac{LISA}, we rigorously constrain the parameter space of $\mathcal{P}_{\zeta}(k)$ and $f_{\mathrm{NL}}$. Furthermore, we examine the effects of different shapes of the primordial power spectrum on these constraints and comprehensively calculate the Bayesian factors for various models. Our results indicate that \acp{SIGW} generated by a monochromatic primordial power spectrum are more likely to dominate current \ac{PTA} observations, with the corresponding constraint on the primordial non-Gaussian parameter being $-10.0<f_{\mathrm{NL}}<1.2$.
\end{abstract}

\maketitle
%%%%%%%%%%%%%%%%%%%%%%%%%%%%%%%%%%%%%%%%
%%%%%%%%%%%%%%%%%%%%%%%%%%%%%%%%%%%%%%%
\acresetall

\section{Introductions}\label{sec:1.0}
Our universe originated from primordial perturbations generated during the inflationary era \cite{Lyth:2005fi,Weinberg:2005vy,Bassett:2005xm}. These primordial perturbations carry the physical information from the inflationary period, influencing all subsequent evolutionary processes of the universe \cite{Baumann:2009ds,Riotto:2002yw,Goldwirth:1991rj,Malik:2008im,Baumann:2018muz}. Through current cosmological observations on different scales, we can determine the physical properties of primordial perturbations on various scales, providing us with crucial information on new physics during the evolution of the universe \cite{Irastorza:2018dyq,Cai:2015emx,Marsh:2015xka,DeFelice:2010aj,Bojowald:2005epg,Gasperini:2002bn,Lyth:1998xn,Kim:2008hd}.

Over the past few decades, cosmological research has made significant strides \cite{Abdalla:2022yfr,ParticleDataGroup:2022pth,ParticleDataGroup:2024cfk}. The cosmological observations, such as \ac{CMB}, \ac{LSS}, \ac{BBN}, and \ac{BAO}, provide us with a wealth of precise cosmological data, enabling us to accurately determine the physical properties of primordial perturbations \cite{Bernardeau:2001qr,Bartolo:2004if,Planck:2013wtn,Planck:2015zfm,eBOSS:2015jyv,Planck:2018nkj,Achucarro:2010da,Stahl:2022did}. Specifically, on large scales ($\gtrsim$1 Mpc), the power spectrum of primordial curvature perturbations is approximately a scale-invariant spectrum, with an amplitude $A_{\zeta} \approx 2\times 10^{-9}$ \cite{Planck:2018vyg}. And the current constraint on the tensor-to-scalar ratio $r$ is $r<0.064$  at $95\%$  confidence level  \cite{Planck:2018jri}. Furthermore, as one of the most important properties of primordial perturbations, the non-Gaussianity of the primordial power spectrum on large scales has also been tightly constrained. For instance, the current cosmological observations reveal that the parameter of local-type primordial non-Gaussianity  $f_{\mathrm{NL}}=-0.9\pm 5.1$ at $68\%$  confidence level \cite{Planck:2019kim}. However, unlike large-scale primordial perturbations, there are no strict observational constraints on small-scale ($\lesssim$1 Mpc) primordial perturbations \cite{Bringmann:2011ut}. Exploring how to leverage different cosmological observations to constrain the small-scale primordial power spectrum and its associated non-Gaussianity represents one of the foremost challenges in further cosmological research.

Various inflationary models can generate primordial curvature perturbations with significant amplitudes and non-Gaussianity on small scales. Examples include multi-field inflation models  \cite{Kristiano:2022maq,Ballesteros:2017fsr,Braglia:2020eai,Palma:2020ejf,Ballesteros:2020qam,Atal:2019erb} and inflation models based on modified gravity theories \cite{Pi:2017gih,Kawai:2021edk,Lin:2020goi,Arya:2023pod,Bamba:2015uma,Chen:2025wcw,Peng:2022ttg}.  To constrain the primordial curvature perturbations and associated non-Gaussianity on small scales, it is essential to consider the influence of these perturbations on small-scale cosmological observations. More precisely, after inflation ends and these large amplitude primordial perturbations re-enter the horizon, they will cause substantial density perturbations that collapse to form \acp{PBH} \cite{Saito:2008jc,Khlopov:2008qy,Carr:2016drx,Sasaki:2018dmp,Carr:2020xqk,DeLuca:2020agl,Musco:2018rwt,Carr:2023tpt,Carr:2021bzv,Liu:2021svg,Choudhury:2023rks,Gouttenoire:2023naa,Belotsky:2014kca}. This process will also inevitably generate \acp{SIGW} \cite{Domenech:2021ztg,Mollerach:2003nq,Ananda:2006af,Baumann:2007zm,Unal:2018yaa,Garcia-Bellido:2017aan}. By analyzing the abundance of \acp{PBH} and the observations of \acp{SIGW} such as \ac{LISA} and \ac{PTA}, we can identify the parameter space for the small-scale primordial power spectrum $\mathcal{P}_{\zeta}\left( k \right)$ and the associated parameter of local-type primordial non-Gaussianity $f_{\mathrm{NL}}$ allowed by current observations.

In this paper, we concentrate on the constraints imposed by current cosmological observations on small-scale local-type primordial non-Gaussianity. More precisely, we consider the following observational constraints:
\tcbset{colback=gray!20, colframe=gray!20, boxrule=0.5mm, arc=0mm, auto outer arc, width=\linewidth} 
\begin{tcolorbox} 
\noindent
\textbf{ 1, \ac{PTA} observations \cite{NANOGrav:2023gor,Reardon:2023gzh,EPTA:2023fyk,Xu:2023wog,NANOGrav:2023hvm,Figueroa:2023zhu,Ellis:2023oxs,Wang:2025kbj,Cang:2023ysz,Tagliazucchi:2023dai}: we assume that \acp{SIGW} dominate the current \ac{PTA} observations and utilize the \ac{PTA} data to perform the Bayesian analysis on the parameter space of $\mathcal{P}_{\zeta}(k)$ and $f_{\mathrm{NL}}$. By combining the model where \acp{SMBHB} dominate \ac{PTA} observations, we calculate the Bayes factors between different models, thereby rigorously analyzing the feasibility of each model in dominating \ac{PTA} observations.}

\

\noindent
\textbf{ 2, Large-scale cosmological observations \cite{Zhou:2024yke,Cang:2022jyc,Ben-Dayan:2019gll,Wang:2023sij}: large-scale cosmological observations such as \ac{CMB}, \ac{BAO}, and \ac{BBN} have provided upper limit on the energy density of \acp{SIGW} $\rho_{\mathrm{GW}}$. Using this upper limit, we can constrain the parameter space of $\mathcal{P}_{\zeta}(k)$ and $f_{\mathrm{NL}}$.}

\

\noindent
\textbf{ 3, \ac{LISA} \cite{LISACosmologyWorkingGroup:2022jok,Iacconi:2024hmg,Flauger:2020qyi}: In scenarios where different models dominate \ac{PTA} observations, we rigorously analyze the \ac{SNR} of \ac{LISA}. This allows us to determine whether the high-frequency regions of the corresponding energy density spectra of \acp{GW} can be observed in the \ac{LISA} band. }

\

\noindent
\textbf{ 4, Abundance of \acp{PBH} \cite{Carr:2020gox}: Given the specific form of  $\mathcal{P}_{\zeta}(k)$ and $f_{\mathrm{NL}}$, we can calculate the abundance of primordial black holes $f_{\mathrm{PBH}}$ and require $f_{\mathrm{PBH}}<1$, thus constraining the parameter space of $\mathcal{P}_{\zeta}(k)$ and $f_{\mathrm{NL}}$.}
\end{tcolorbox}
\noindent
Using the aforementioned four types of cosmological observations, we can rigorously quantify the constraints on the small-scale primordial power spectrum and primordial non-Gaussianity.

This paper is organized as follows. In Sec.~\ref{sec:2.0}, we study the energy density spectra of \acp{SIGW} and analyze the constraints on $\mathcal{P}_{\zeta}(k)$ and $f_{\mathrm{NL}}$ imposed by \ac{PTA}, \ac{LISA}, and large-scale cosmological observations. In Sec.~\ref{sec:3.0}, we calculate the abundance of \acp{PBH} and examine the impact of \ac{PBH} abundance on the constraints of $f_{\mathrm{NL}}$. In Sec.~\ref{sec:4.0}, we investigate the impact of various forms of the primordial power spectrum on the constraints of $f_{\mathrm{NL}}$. Additionally, we compute the Bayes factors between different models using the current \ac{PTA} data. Finally, we summarize our results and give some discussions in Sec.~\ref{sec:5.0}.

\section{Scalar induced gravitational waves}\label{sec:2.0}
In June 2023, the \ac{PTA} collaborations NANOGrav \cite{NANOGrav:2023gor}, EPTA \cite{EPTA:2023fyk}, PPTA \cite{Reardon:2023gzh}, and the CPTA \cite{Xu:2023wog} have reported positive evidence for an isotropic, stochastic background of \acp{GW} within the nHz frequency range. As mentioned in ref.~\cite{NANOGrav:2023hvm}, among the numerous potential contributors to the \ac{SGWB}, \acp{SIGW} show the highest Bayesian factor, which makes them one of the most likely dominant sources. In this case, considering the scenario where \acp{SIGW} dominate the current nHz frequency band of the \ac{SGWB}, we perform Bayesian analysis on the current \ac{PTA} observational data. This allows us to determine the current \ac{PTA} constraints on small-scale  primordial power spectrum $\mathcal{P}_{\zeta}\left(k \right)$ and the corresponding parameter of local-type primordial non-Gaussianity $f_{\mathrm{NL}}$ \cite{Yi:2023npi,Wang:2023ost,Perna:2024ehx,Papanikolaou:2024kjb,Wang:2024rdf,Kristiano:2021urj,He:2024luf}. 

In the case of the local-type primordial non-Gaussianity, the non-Gaussian primordial curvature perturbation can be expressed as a local perturbative expansion around the Gaussian primordial curvature perturbation. In momentum space, the primordial curvature perturbation can be written as \cite{Cai:2018dig,Domenech:2021ztg}
\begin{equation}
	\begin{aligned}	\zeta^{\mathrm{ng}}_{\mathbf{k}}=\zeta^{\mathrm{g}}_{\mathbf{k}}+\frac{3}{5}f_{\mathrm{NL}}\int\frac{{\dif}^3n}{(2\pi)^{3/2}} \zeta^{\mathrm{g}}_{\mathbf{k}-\mathbf{n}}\zeta^{\mathrm{g}}_{\mathbf{n}}  \ ,
	\end{aligned}\label{eq:ngk}
\end{equation}
where $\mathbf{n}$ is the three dimensional momentum variable. $\zeta^{\mathrm{g}}_{\mathbf{k}}$ is the Gaussian primordial curvature perturbation. 

The second-order \acp{SIGW} can be expressed as
\begin{equation}
	\begin{aligned}
		h^{\lambda,(2)}_{\mathbf{k}}(\eta)= \int \frac{\dif^3 p}{(2 \pi)^{3 / 2}}  \varepsilon^{\lambda, l m}(\mathbf{k})p_l p_mI^{(2)} \zeta_{\mathbf{k}-\mathbf{p}}  \zeta_{\mathbf{p}} \ ,
	\end{aligned}\label{eq:2h}
\end{equation}
where $\varepsilon^{\lambda, l m}(\mathbf{k})$$(\lambda=+,\times)$ are polarization tensors. The explicit expressions of the second-order kernel functions $I^{(2)}$ can be found in Ref.~\cite{Kohri:2018awv}. By analyzing the two-point correlation function of second-order \acp{SIGW}, we can determine the corresponding energy density spectrum. The calculation of two-point correlation function $\langle h^{\lambda,(2)}_{\mathbf{k}} h^{\lambda',(2)}_{\mathbf{k}'}  \rangle$ involves the four-point correlation function associated with primordial curvature perturbations: $\langle\zeta_{\mathbf{k}-\mathbf{p}}\zeta_{\mathbf{p}}\zeta_{\mathbf{k}'-\mathbf{p}'}\zeta_{\mathbf{p}'}\rangle$. Combining the expression of local-type non-Gaussian primordial curvature perturbations provided in Eq.~(\ref{eq:ngk}), the total  energy density spectrum of second-order \acp{SIGW} can be written as \cite{Adshead:2021hnm,Li:2025met}
\begin{equation}\label{eq:Ome}
\begin{aligned}	
\bar{\Omega}^{(2)}_{\mathrm{GW}}(k)&=\bar{\Omega}_{\mathrm{GW}}^G+\bar{\Omega}_{\mathrm{GW}}^H+\bar{\Omega}_{\mathrm{GW}}^C+\bar{\Omega}_{\mathrm{GW}}^Z \\
&+\bar{\Omega}_{\mathrm{GW}}^R+\bar{\Omega}_{\mathrm{GW}}^P+\bar{\Omega}_{\mathrm{GW}}^N \ , 
\end{aligned} 
\end{equation}
where the energy density spectrum of the second-order \acp{SIGW} is categorized into seven distinct loop diagram contributions. In Eq.~(\ref{eq:Ome}), the contributions of Gaussian one-loop diagrams, proportional to $A_{\zeta}^2$, are denoted by $\bar{\Omega}_{\mathrm{GW}}^G$. The 
symbols $\bar{\Omega}_{\mathrm{GW}}^H$, $\bar{\Omega}_{\mathrm{GW}}^C$, and  $\bar{\Omega}_{\mathrm{GW}}^Z$ denote non-Gaussian contributions of two-loop diagrams proportional to $A_{\zeta}^3\left(f_{\mathrm{NL}}\right)^2$. The 
symbols $\bar{\Omega}_{\mathrm{GW}}^R$, $\bar{\Omega}_{\mathrm{GW}}^P$, and $\bar{\Omega}_{\mathrm{GW}}^N$ correspond to non-Gaussian contributions of three-loop diagrams  proportional to $A_{\zeta}^4\left(f_{\mathrm{NL}}\right)^4$. The explicit expressions of these loop diagrams of second-order \acp{SIGW} in Eq.~(\ref{eq:Ome}) can be found in Ref.~\cite{Li:2023qua}.  

Furthermore, due to the significant primordial perturbations on small scales, higher-order cosmological perturbations significantly affect the total energy density spectrum of \acp{SIGW}, thus modifying the parameter space of $\mathcal{P}_{\zeta}(k)$ and $f_{\mathrm{NL}}$  determined by \acp{PTA} observations. Specifically, the third-order \acp{SIGW} can be expressed as \cite{Zhou:2021vcw,Chang:2022nzu}
\begin{eqnarray}
h^{\lambda,(3)}_{\mathbf{k}}(\eta)&=&\sum_{i=1}^{4}h_{i,\mathbf{k}}^{\lambda,(3)}(\eta)=\int \frac{\dif^3 p}{(2 \pi)^{3 / 2}} \int \frac{\dif^3 q}{(2 \pi)^{3 / 2}}  \nonumber\\
&\times&\varepsilon^{\lambda, l m}(\mathbf{k})\mathbb{P}^{i}_{lm} I_{i}^{(3)} \zeta_{\mathbf{k}-\mathbf{p}} \zeta_{\mathbf{p}-\mathbf{q}} \zeta_{\mathbf{q}} \ , \label{eq:3h}
\end{eqnarray}
where the momentum polynomials $\mathbb{P}^{i}_{lm}$$(i=1,2,3,4)$ are given by
\begin{eqnarray}
\mathbb{P}^{1}_{lm}&=&q_m \left(p_l-q_l\right) \ , \ \mathbb{P}^{2}_{lm}=\Lambda_{l m}^{r s}(\mathbf{p}) q_r q_s \ , \label{eq:3h1}\\
\mathbb{P}^{3}_{lm}&=&\left(\mathcal{T}_{m}^r(\mathbf{p}) p_{l}+\mathcal{T}_{l}^r(\mathbf{p}) p_{m}\right)  \frac{p^s}{ p^2} q_r q_s \ , \label{eq:3h3}\\
\mathbb{P}^{4}_{lm}&=&p_l p_m  \  . \label{eq:3h4}
\end{eqnarray}
In Eq.~(\ref{eq:3h1}), $\Lambda_{i j}^{lm}(\mathbf{p})$ represents the transverse and traceless operator, defined as
\begin{eqnarray}\label{eq:LA}
    \Lambda_{i j}^{lm}(\mathbf{p})=\mathcal{T}_{i}^{l}(\mathbf{p}) \mathcal{T}_{j}^{m}(\mathbf{p})-\frac{1}{2} \mathcal{T}_{ij}(\mathbf{p}) \mathcal{T}^{l m}(\mathbf{p})  \ ,
\end{eqnarray}
where $\mathcal{T}_{ij}(\mathbf{p})=\delta_{ij}-p_ip_j/p^2$. In Eq.~(\ref{eq:3h}), the contribution of third-order \acp{SIGW} consists of four parts, denoted as $h_{i,\mathbf{k}}^{\lambda,(3)}$$(i=1,2,3,4)$. They correspond respectively to the third-order gravitational waves directly induced by first-order scalar perturbations, and to the third-order gravitational waves jointly induced by first-order scalar perturbations and three types of second-order perturbations. The explicit expressions of the third-order kernel functions $I^{(3)}_i$$(i=1,2,2,3,4)$ in Eq.~(\ref{eq:3h}) are provided in Ref.~\cite{Zhou:2021vcw}. 

When we consider \acp{SIGW} up to the third order, the two-point correlation function of the gravitational waves can be expressed as
\begin{eqnarray}\label{eq:coh}
   \langle h^{\lambda}_{\mathbf{k}} h^{\lambda'}_{\mathbf{k}'}  \rangle &=& \langle h^{\lambda,(2)}_{\mathbf{k}} h^{\lambda',(2)}_{\mathbf{k}'}  \rangle+ \langle h^{\lambda,(3)}_{\mathbf{k}} h^{\lambda',(3)}_{\mathbf{k}'}  \rangle \nonumber\\
   &+&2~\langle h^{\lambda,(3)}_{\mathbf{k}} h^{\lambda',(2)}_{\mathbf{k}'}  \rangle \ .
\end{eqnarray}
Here, we have ignored the potential large-amplitude primordial gravitational waves on small scales \cite{Wu:2024qdb,Fu:2023aab,Gorji:2023sil}. As previously discussed, in Eq.~(\ref{eq:coh}), $\langle h^{\lambda,(2)}_{\mathbf{k}} h^{\lambda',(2)}_{\mathbf{k}'}  \rangle$ represents the two-point correlation function of second-order \acp{SIGW}, proportional to the four-point correlation function of primordial curvature perturbations: $\langle\zeta_{\mathbf{k}-\mathbf{p}}\zeta_{\mathbf{p}}\zeta_{\mathbf{k}'-\mathbf{p}'}\zeta_{\mathbf{p}'}\rangle$.  Similarly, the two-point correlation function of third-order \acp{SIGW} $\langle h^{\lambda,(3)}_{\mathbf{k}} h^{\lambda',(3)}_{\mathbf{k}'}  \rangle$ is proportional to the six-point correlation function of primordial curvature perturbations: $\langle\zeta_{\mathbf{k}-\mathbf{p}}\zeta_{\mathbf{p}-\mathbf{q}}\zeta_{\mathbf{q}}\zeta_{\mathbf{k}'-\mathbf{p}'}\zeta_{\mathbf{p}'-\mathbf{q}'}\zeta_{\mathbf{q}'}\rangle$. And the cross-correlation function $\langle h^{\lambda,(3)}_{\mathbf{k}} h^{\lambda',(2)}_{\mathbf{k}'}  \rangle$ is proportional to the five-point correlation function of primordial curvature perturbations: $\langle\zeta_{\mathbf{k}-\mathbf{p}}\zeta_{\mathbf{p}-\mathbf{q}}\zeta_{\mathbf{q}}\zeta_{\mathbf{k}'-\mathbf{p}'}\zeta_{\mathbf{p}'}\rangle$. In addition, the cross two-point correlation function $\langle h^{\lambda,(3)}_{\mathbf{k}} h^{\lambda',(2)}_{\mathbf{k}'}  \rangle$ will only modify the total energy density spectrum of \acp{SIGW} if primordial non-Gaussianity is present. The lowest-order contribution of this modification is proportional to $A_{\zeta}^3f_{\mathrm{NL}}$, corresponding to the two-loop diagrams of \acp{SIGW} \cite{Chang:2023aba}. In summary, when considering \acp{SIGW} up to the third order, the corresponding one-loop $(\sim A_{\zeta}^2)$ and two-loop $(\sim A_{\zeta}^3)$ contributions can be expressed as 
\begin{eqnarray}\label{eq:Otot}
\bar{\Omega}^{\mathrm{tot}}_{\mathrm{GW}}(k)&=&\bar{\Omega}_{\mathrm{GW}}^G+\bar{\Omega}_{\mathrm{GW}}^H+\bar{\Omega}_{\mathrm{GW}}^C+\bar{\Omega}_{\mathrm{GW}}^Z \nonumber\\
&+&\bar{\Omega}^{(3)}_{\mathrm{GW}}
+2\bar{\Omega}^{(3,2)}_{\mathrm{GW}} \ .
\end{eqnarray}
{It is important to emphasize that when considering third-order \acp{SIGW}, the influence of the parameter $f_{\mathrm{NL}}$ on the total energy density spectrum $\bar{\Omega}^{\mathrm{tot}}_{\mathrm{GW}}(k)$ in Eq.~(\ref{eq:Otot}) differs significantly from that of second-order \acp{SIGW} in Eq.~(\ref{eq:Ome}). Specifically, the energy density spectrum of second-order \acp{SIGW} $\bar{\Omega}^{(2)}_{\mathrm{GW}}(k)$ in Eq.~(\ref{eq:Ome}) is proportional to $A_{\zeta}^2+(f_{\mathrm{NL}})^2A_{\zeta}^3+(f_{\mathrm{NL}})^4A_{\zeta}^4$. Hence, regardless of whether $f_{\mathrm{NL}}$ is positive or negative, as long as its absolute value remains the same, $\bar{\Omega}^{(2)}_{\mathrm{GW}}(k)$ yields identical results and is always strictly positive. However, for the total energy density spectrum $\bar{\Omega}^{\mathrm{tot}}_{\mathrm{GW}}(k)$ in Eq.~(\ref{eq:Otot}), which includes third-order \acp{SIGW}, the spectrum $\bar{\Omega}^{(3,2)}_{\mathrm{GW}}$ corresponding to the cross-correlation function $\langle h^{\lambda,(2)}_{\mathbf{k}} h^{\lambda',(3)}_{\mathbf{k}'}  \rangle$ is proportional to $A_{\zeta}^3f_{\mathrm{NL}}$. Therefore, even if the absolute value of $f_{\mathrm{NL}}$ remains unchanged, altering the sign of $f_{\mathrm{NL}}$ can still affect $\bar{\Omega}^{(3,2)}_{\mathrm{GW}}$ and $\bar{\Omega}^{\mathrm{tot}}_{\mathrm{GW}}(k)$. In particular, when $f_{\mathrm{NL}}<0$, the perturbative expansion coefficient in front of $\bar{\Omega}^{(3,2)}_{\mathrm{GW}}$ becomes negative; as a result, the contributions from $\bar{\Omega}^{(3,2)}_{\mathrm{GW}}$ suppress the total energy density spectrum of \acp{SIGW}.} The explicit expressions of third-order corrections $\bar{\Omega}^{(3)}_{\mathrm{GW}}$ and $\bar{\Omega}^{(3,2)}_{\mathrm{GW}}$ can be found in Refs.~\cite{Chang:2023aba,Chang:2023vjk}. 

In this study, we focus on the following two types of energy density spectra:

\noindent
\textbf{1.} The total energy density spectrum of second-order \acp{SIGW} as described in Eq.~(\ref{eq:Ome}). 

\noindent
\textbf{2.} The complete one-loop and two-loop contributions involving up to third-order \acp{SIGW} as presented in Eq.~(\ref{eq:Otot}). 

\noindent
Furthermore, Eq.~(\ref{eq:Ome}) and Eq.~(\ref{eq:Otot}) present the formulas for the energy density spectrum of \acp{SIGW} during the \ac{RD} era. Taking into account the thermal history of the universe, we obtain the current energy density spectrum \cite{Wang:2019kaf}
\begin{equation}
    \bar{\Omega}_{\mathrm{GW,0}}(k) = \Omega_{\mathrm{rad},0}\left(\frac{g_{*,\rho,e}}{g_{*,\rho,0}}\right)\left(\frac{g_{*,s,0}}{g_{*,s,e}}\right)^{4/3}\bar{\Omega}_{\mathrm{GW}}(k) \ ,
\end{equation}
where $\Omega_{\mathrm{rad},0}$ ($ =4.2\times 10^{-5}h^{-2}$) is the energy density fraction of radiations today. The effect numbers of relativistic species $g_{*,\rho}$ and $g_{*,s}$ can be found in Ref.~\cite{Saikawa:2018rcs}. And  the dimensionless Hubble constant is $h = 0.6736$ \cite{Planck:2018vyg}.

\subsection{PTA observations}\label{sec:2.1}
Given the specific form of the primordial power spectrum $\mathcal{P}_{\zeta}(k)$, we can calculate the energy density spectrum of \acp{SIGW}. In this section, we consider the \ac{LN} primordial power spectrum \cite{Pi:2020otn}
\begin{equation} \label{eq:LN}
\mathcal{P}^{\mathrm{LN}}_\zeta (k)=\frac{A_{\zeta}}{\sqrt{2 \pi \sigma^2} } \exp \left(-\frac{1}{2 \sigma^2} \ln ^2\left(k / k_*\right)\right) \ ,
\end{equation}
where $A_{\zeta}$ is the amplitude of primordial power spectrum and $k_*=2\pi f_*$ is the wavenumber at which the primordial power spectrum has a \ac{LN} peak. And the parameter $\sigma$ indicates the width of the \ac{LN} primordial power spectrum. 

To characterize the parameter space of $\mathcal{P}^{\mathrm{LN}}_{\zeta}(k)$ and $f_{\mathrm{NL}}$ inferred from \ac{PTA} observations, we construct the likelihood function using \ac{KDE} representations of the free spectra \cite{Mitridate:2023oar,Lamb:2023jls,Moore:2021ibq}. The likelihood is given by
\begin{equation} \label{eq:likelihood}
    \ln \mathcal{L}(d|\theta) = \sum_{i=1}^{N_f} p(\Phi_i,\theta)\ .
\end{equation}
Here $p(\Phi_i,\theta)$ represents the probability of $\Phi_i$ given the parameter $\theta$, and $\Phi_i = \Phi(f_i)$ denotes the time delay
\begin{equation} \label{eq:timedelay}
    \Phi(f) = \sqrt{\frac{H_0^2 \Omega_{\mathrm{GW}}(f)}{8\pi^2 f^5 T_{\mathrm{obs}}}} \ ,
\end{equation}
where $H_0=h\times 100 \mathrm{km/s/Mpc}$ is the present-day value of the Hubble constant. We directly use the \acp{KDE} representation of the first 14 frequency HD-correlated free spectrum in NANOGrav 15-year dataset \cite{Nanograv:KDE}. Bayesian analysis is performed by \textsc{bilby} \cite{bilby_paper} with its built-in \textsc{dynesty} nested sampler \cite{Speagle:2019ivv,dynesty_software}. Furthermore, to investigate the impact of astrophysical sources of the \ac{SGWB} on current \ac{PTA} observations, we consider the \ac{SGWB} generated by \acp{SMBHB}, with the corresponding energy density spectrum given by \cite{Mitridate:2023oar,NANOGrav:2023hvm}
\begin{equation} \label{eq:SMBHB}
    \Omega^{\mathrm{BH}}_{\mathrm{GW}}(f) = \frac{2\pi^2 A_{\mathrm{BHB}}^2}{3H_0^2 h^2} (\frac{f}{\mathrm{year}^{-1}})^{5-\gamma_{\mathrm{BHB}}}\mathrm{year}^{-2} \ .
\end{equation}
Here, the prior distribution for $(\log_{10}A_{\mathrm{BHB}}, \gamma_{\mathrm{BHB}})$ follows a multivariate normal distribution \cite{NANOGrav:2023hvm}, whose mean and covariance matrix are given by
\begin{equation} \label{eq:prior_SMBHB}
    \boldsymbol{\mu}_{\mathrm{BHB}} =\begin{pmatrix} -15.6
 \\ 4.7 \end{pmatrix} , 
\boldsymbol{\sigma}_{\mathrm{BHB}}=0.1\begin{pmatrix}
2.8  & -0.026\\
-0.026  & 2.8
\end{pmatrix} \ .
\end{equation}

\begin{figure}[htbp]
    \captionsetup{
      justification=raggedright,
      singlelinecheck=true
    }
    \centering
	\subfloat[Posterior distributions of $\Omega^{\mathrm{BH}}_{\mathrm{GW}}(k)+\bar{\Omega}^{(2)}_{\mathrm{GW}}(k)$ and $\bar{\Omega}^{(2)}_{\mathrm{GW}}(k)$.\label{fig:corner_eq3}]  {\includegraphics[width=\columnwidth]{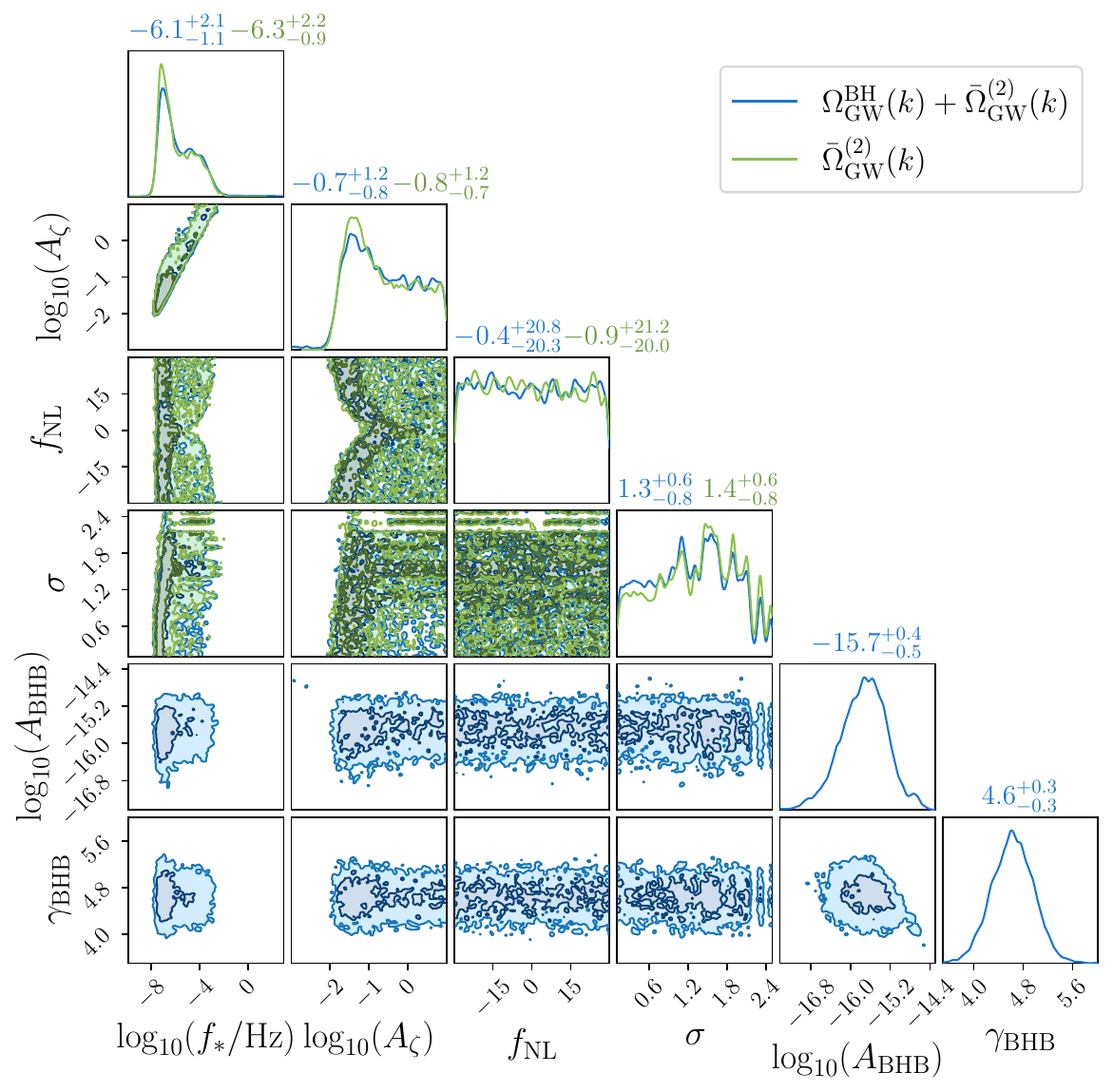}}
    \hspace{0.05cm}
        \subfloat[Posterior distributions of $\bar{\Omega}^{(2)}_{\mathrm{GW}}(k)$ and $\bar{\Omega}^{\mathrm{tot}}_{\mathrm{GW}}(k)$.\label{fig:corner_eq3vseq10}]{\includegraphics[width=\columnwidth]{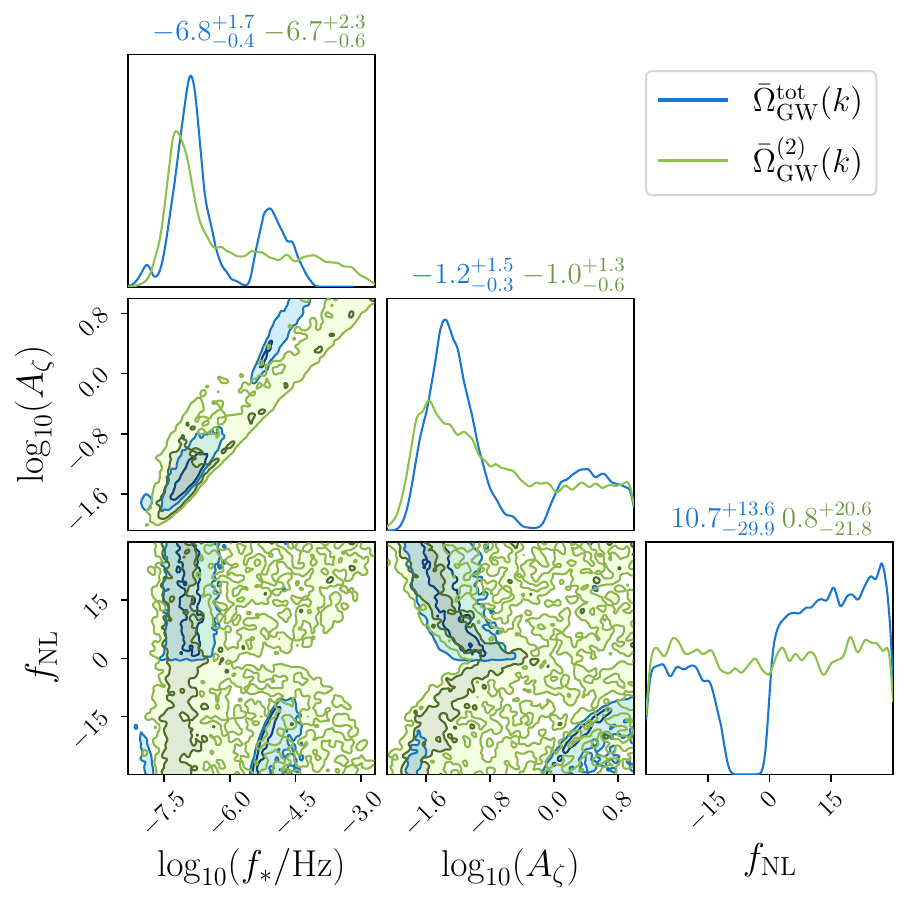}} 
\caption{\label{fig:eq3eq10} The corner plot of the posterior distributions. The contours in the off-diagonal panels denote the $68\% $ and $95 \%$ credible intervals of the 2D posteriors. The numbers above the figures represent the median values and $1$-$\sigma$ ranges of the parameters.  \textbf{(a): }The blue and green solid curves correspond to the second-order \ac{SIGW} energy spectrum in Eq.~(\ref{eq:Ome}) with or without \ac{SMBHB}. \textbf{(b): } The green and blue lines represent the energy density spectra of second-order \acp{SIGW} in Eq.~(\ref{eq:Ome}) and the complete one-loop and two-loop energy spectra incorporating up to third-order \acp{SIGW} in Eq.~(\ref{eq:Otot}), respectively. We have fixed $\sigma=1$.}
\end{figure}

\begin{figure}[!ht]
\centering
\includegraphics[width=\linewidth]{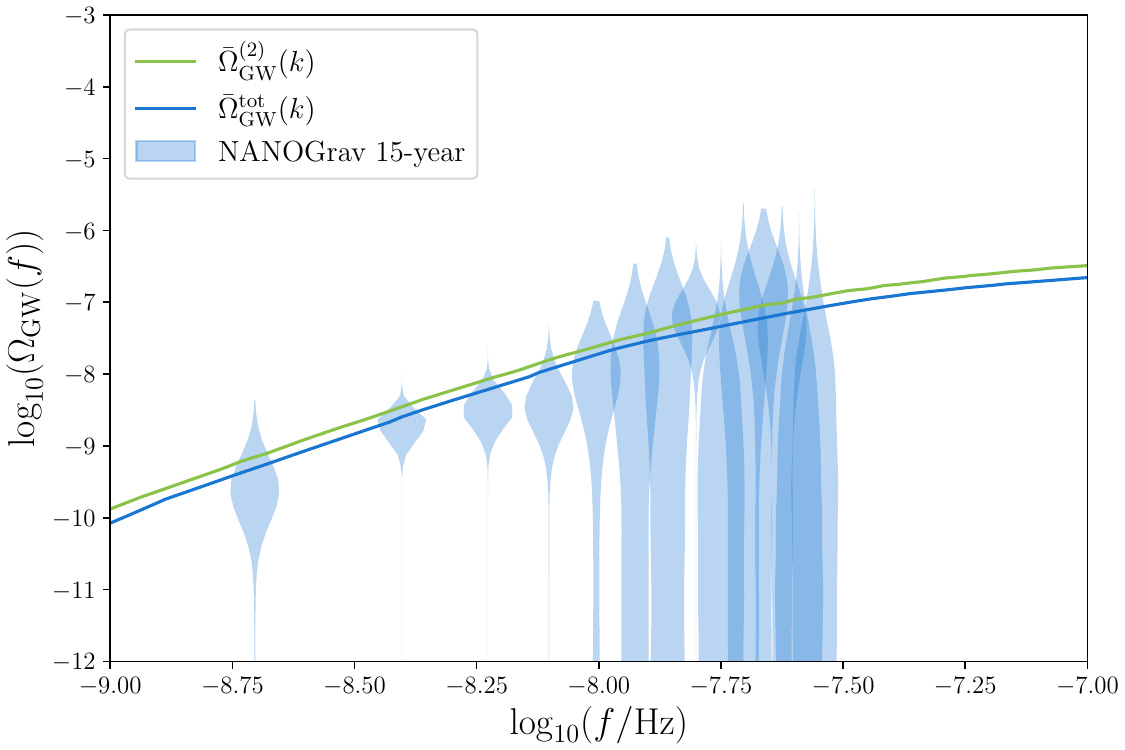}
\caption{The energy density spectra $\bar{\Omega}^{(2)}_{\mathrm{GW}}(k)$ and $\bar{\Omega}^{\mathrm{tot}}_{\mathrm{GW}}(k)$ correspond to Fig.~\ref{fig:corner_eq3vseq10}. The energy density spectra derived from the free spectrum of the NANOGrav 15-year are shown with blue. The green curves represent the energy density spectra of \acp{GW} with different line styles labeled in the figure. Specifically, the parameters for the blue solid line are: $\log_{10}(A_{\zeta})=-1.51$,  $\log_{10}(f_*/\mathrm{Hz})=-7.14$,  and $f_{\mathrm{NL}}=22.3$; and for the green solid line, the parameters are: $\log_{10}(A_{\zeta})=-1.32$,  $\log_{10}(f_*/\mathrm{Hz})=-6.991$, and $f_{\mathrm{NL}}=23.4$. These parameters are selected based on the \acf{MAP} estimate for each parameter. } \label{fig:violinplot_map}
\end{figure}

For the second-order \acp{SIGW} in Eq.~(\ref{eq:Ome}), the posterior distributions are depicted in Fig.~\ref{fig:corner_eq3}, where the prior distributions of $\log(f_*/\mathrm{Hz})$, $\log(A_\zeta)$, $\sigma$, and $f_{\mathrm{NL}}$ are uniformly distributed over the ranges $[-10,3]$, $[-3,1]$, $[0.1,2.5]$, and $[-30,30]$ respectively. The corresponding energy density spectra of second-order \acp{SIGW} are given in Fig.~\ref{fig:violinplot_map}. Furthermore, to determine the impact of third-order \acp{SIGW} on the amplitude of the primordial power spectrum $A_{\zeta}$, we calculate the energy density spectrum in Eq.~(\ref{eq:Ome}) and Eq.~(\ref{eq:Otot}), with the parameter $\sigma$ fixed. Fig.~\ref{fig:corner_eq3vseq10} compares the corresponding posterior distributions, showing that the presence of third-order \acp{SIGW} significantly modifies the parameter space of $A_{\zeta}$ and $f_{\mathrm{NL}}$ determined by current \ac{PTA} observations. More precisely,  after taking into account the contributions of the third-order \acp{SIGW}, the blue curve of $f_{\mathrm{NL}}$ in Fig.~\ref{fig:corner_eq3vseq10} is no longer symmetric, and the parameter interval $f_{\mathrm{NL}}\in [-10,-1]$ is significantly excluded. When $f_{\mathrm{NL}}\in [-10,-1]$, the cross-correlation function $\langle h^{\lambda,(3)}_{\mathbf{k}}(\eta) h^{\lambda',(2)}_{\mathbf{k}'}(\eta) \rangle$ will suppress the total energy density spectrum, which prevents the total energy density spectrum of \acp{SIGW} from fitting well with the observational data of \ac{PTA} in this parameter interval.

\subsection{SNR of LISA}\label{sec:2.2}
In scenarios where the high-frequency region of the energy spectrum of \acp{SIGW} is within detection range of \ac{LISA}, we can combine \ac{PTA} observational data to thoroughly analyze the impact of $\mathcal{P}_{\zeta}\left(k \right)$ and $f_{\mathrm{NL}}$ on the \ac{SNR} of \ac{LISA}, thereby determining the influence of the high-frequency \acp{SIGW} on LISA observations. The \ac{SNR} $\rho$ of \ac{LISA} is given by \cite{Siemens:2013zla}
\begin{equation}
    \rho = \sqrt{T}\left[ \int \dif f\left(\frac{\bar{\Omega}_{\mathrm{GW},0}(f)}{\Omega_\mathrm{n}(f)}\right)^2\right]^{1/2} \ ,
\end{equation}
where $T$ is the observation time and we set $T=4$ years here. $\Omega_\mathrm{n}(f)=2\pi^2f^3S_n/3H_0^2$, where $H_0$ is the Hubble constant and $S_n$ is the strain noise power spectral density \cite{Robson:2018ifk}. 

\begin{figure}[htbp]
    \centering
    \includegraphics[width=\columnwidth]{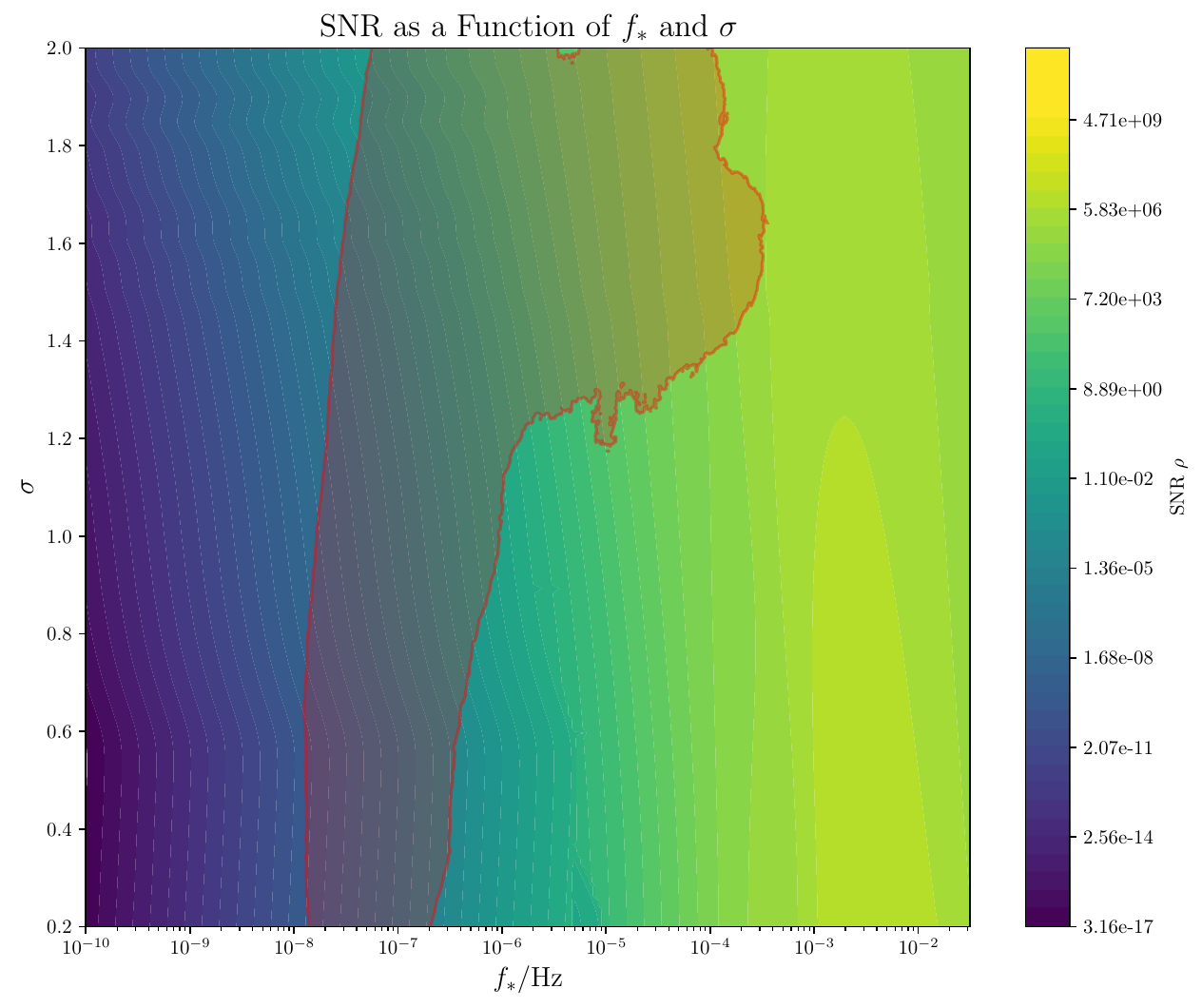}   
\caption{\label{fig:heatmap_fsigma}The \ac{SNR} of \ac{LISA} as a function of $f_*$ and $\sigma$. The heatmap shows the \ac{SNR} for second-order \ac{SIGW} energy spectrum $\bar{\Omega}^{(2)}_{\mathrm{GW}}(k)$ with fixed values of $A_\zeta = 0.1$ and $f_{\mathrm{NL}} = 5$. The red shading represents the $68\%$ credible interval corresponding to the two-dimensional posterior distribution of $\bar{\Omega}^{(2)}_{\mathrm{GW}}(k)$ shown in Fig.~\ref{fig:corner_eq3}, obtained using the \ac{KDE} method implemented in ChainConsumer \cite{Hinton2016,Hinton2016_software}.}
\end{figure}

As shown in Fig.~\ref{fig:heatmap_fsigma}, we provide a two-dimensional distribution plot of \ac{SNR} of \ac{LISA} for $\sigma$ and $f_*$. The results indicate that when \acp{SIGW} dominate the current \ac{PTA} observations, the corresponding energy density spectrum in the high-frequency region will influence observations in the \ac{LISA} frequency band. Moreover, as elaborated in Sec.~\ref{sec:3.2}, large-amplitude primordial perturbations at small scales can lead to the formation of primordial black holes, while binary \ac{PBH} mergers also contribute to the \ac{SGWB}. When \acp{SIGW} dominate \ac{PTA} observations, the \ac{SGWB} produced by binary \ac{PBH} mergers can also impact \ac{GW} observations at higher frequencies.

\subsection{Constraints from large scales cosmological observations}\label{sec:2.3}
\begin{figure}[htbp]
    \centering
    \includegraphics[width=.9\columnwidth]{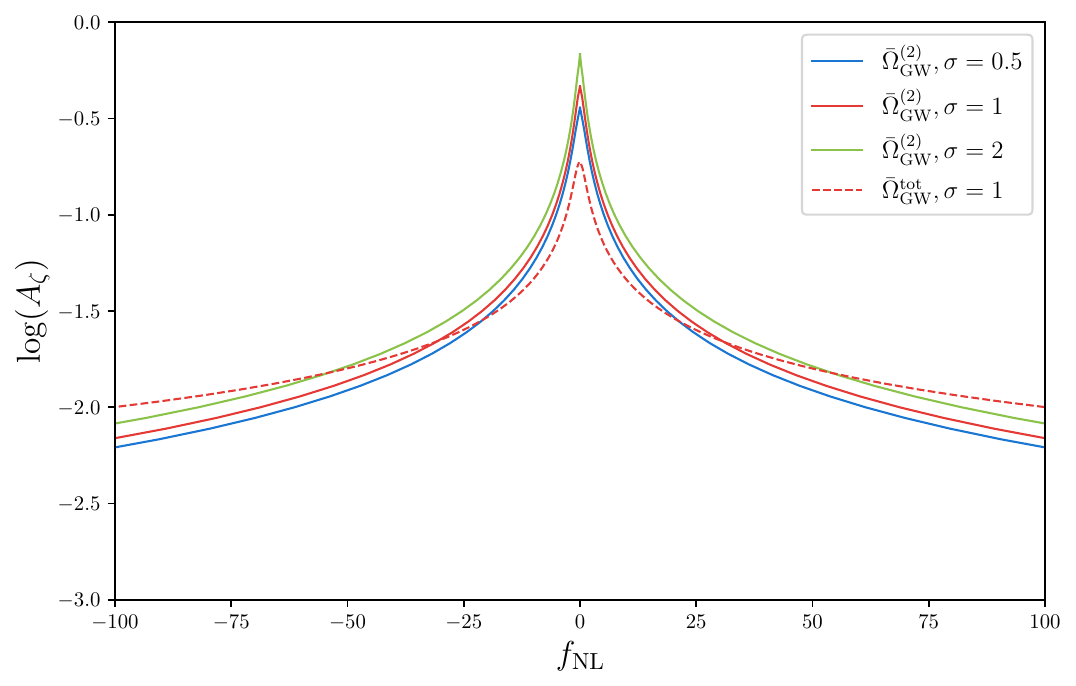}
\caption{\label{fig:cosmo} The upper limits of the amplitude of \ac{LN} primordial power spectrum  $A_\zeta$
 as determined by large-scale observations under different non-Gaussian parameter $f_{\mathrm{NL}}$.  The figure illustrates the upper limits of $A_{\zeta}$, as determined by the energy density spectra of Eq.~(\ref{eq:Ome}) (solid line) and Eq.~(\ref{eq:Otot}) (dashed line). }
\end{figure}

Besides the direct observation of the energy density spectrum of \acp{SIGW}, \acp{SIGW} can serve as an additional radiation component, affecting the large-scales cosmological observations \cite{Zhou:2024yke,Wright:2024awr,Ben-Dayan:2019gll}. More precisely, the total energy density of \acp{SIGW} satisfies
\begin{equation}\label{eq:rhup}
h^2\rho_{\mathrm{GW}}=\int_{f_{\mathrm{min}}}^{\infty} h^2\Omega_{\mathrm{GW},0}(k) \dif\left(\ln k\right) < 2.9\times 10^{-7} 
\end{equation}
at $95\%$ confidence level for \ac{CMB}$+$\ac{BAO} data \cite{Clarke:2020bil}. It should be noted that the large-scale cosmological observation constraints in Ref.~\cite{Clarke:2020bil} are stronger than the constraints obtained from the relativistic degrees of freedom $N_{\text{eff}}$ in Refs.~\cite{Wang:2023sij,Zhou:2024yke}. More precisely, taking into account only the constraints from $N_{\text{eff}}$, the energy density spectrum of \acp{SIGW} satisfies $h^2\rho_{\mathrm{GW}}<2.11\times 10^{-6}$ at the $95\%$ confidence level, which represents a weaker constraint compared to that provided in Eq.~(\ref{eq:rhup}). In Fig.~\ref{fig:cosmo}, we present the parameter space of $\mathcal{P}_{\zeta}(k)$ and $f_{\mathrm{NL}}$ determined by Eq.~(\ref{eq:rhup}).

{As depicted in Fig.~\ref{fig:cosmo}, the peak values of all curves are located near zero, indicating that when the non-Gaussianity parameter $f_{\mathrm{NL}}=0$, the constraints from current large-scale cosmological observations on the amplitude of the primordial power spectrum $A_{\zeta}$ are weakest. To understand this behavior, it is necessary to examine the dependence of the energy density spectrum of \acp{SIGW} on parameters $A_{\zeta}$ and $f_{\mathrm{NL}}$. Specifically, as indicated by Eq.~(\ref{eq:rhup}), current large-scale cosmological observations impose an upper bound on the energy density of \acp{SIGW}. When $f_{\mathrm{NL}}=0$, the energy density spectra in Eq.~(\ref{eq:Ome}) and Eq.~(\ref{eq:Otot}) can be expressed as
\begin{eqnarray}\label{eq:duibi}
    \bar{\Omega}^{(2)}_{\mathrm{GW}}(k)&=&\bar{\Omega}_{\mathrm{GW}}^G(k) \ , \ \bar{\Omega}^{\mathrm{tot}}_{\mathrm{GW}}(k)=\bar{\Omega}_{\mathrm{GW}}^G+\bar{\Omega}^{(3)}_{\mathrm{GW}} \ , \nonumber\\
    &~& f_{\mathrm{NL}}= 0 \ ,
\end{eqnarray}
where $\bar{\Omega}_{\mathrm{GW}}^G$ and $\bar{\Omega}^{(3)}_{\mathrm{GW}}$ represent the energy density spectra of second-order and third-order \acp{SIGW} under the assumption of Gaussian primordial curvature perturbations. Here, $\bar{\Omega}_{\mathrm{GW}}^G$ and $\bar{\Omega}^{(3)}_{\mathrm{GW}}$ are, respectively, proportional to $A_{\zeta}^2$ and $A_{\zeta}^3$. Applying the inequality in Eq.~(\ref{eq:rhup}), constraints can be placed on the amplitude of the primordial power spectrum $A_{\zeta}$. When $f_{\mathrm{NL}}\ne 0$, comparison among Eq.~(\ref{eq:duibi}), Eq.~(\ref{eq:Ome}), and Eq.~(\ref{eq:Otot}) reveals that the presence of primordial non-Gaussianity introduces additional contributions to the energy density spectrum of \acp{SIGW}. Consequently, to satisfy the constraint in Eq.~(\ref{eq:rhup}) when $f_{\mathrm{NL}}\ne 0$, the amplitude of the primordial power spectrum $A_{\zeta}$ must be smaller. This explains the origin of the peak positions observed in Fig.~\ref{fig:cosmo}.} In addition, the total energy density spectrum of second-order \acp{SIGW} $\Omega^{(2)}_{\mathrm{GW}}$
 in Eq.~(\ref{eq:Ome}) is proportional to $A_{\zeta}^2+(f_{\mathrm{NL}})^2A_{\zeta}^3+(f_{\mathrm{NL}})^4A_{\zeta}^4$
, creating a degeneracy for opposite values of $f_{\mathrm{NL}}$
, where the upper limits of $A_{\zeta}$ for $f_{\mathrm{NL}}$ and $-f_{\mathrm{NL}}$ are identical. However, in Eq.~(\ref{eq:Otot}), where one-loop and two-loop contributions extend to third order, the corresponding energy density spectrum $\Omega^{\mathrm{tot}}_{\mathrm{GW}}$ is proportional to $A_{\zeta}^2+A_{\zeta}^3+f_{\mathrm{NL}}A_{\zeta}^3+(f_{\mathrm{NL}})^2A_{\zeta}^3$, causing the upper limits of $A_{\zeta}$ to differ for parameters $f_{\mathrm{NL}}$ with equal magnitudes but opposite signs. Furthermore, when $|f_{\mathrm{NL}}|$ is small, the third-order \acp{SIGW} contribution from Eq.~(\ref{eq:Otot}) suppresses the upper bound of $A_{\zeta}$. As $|f_{\mathrm{NL}}|$ increases, the three-loop contribution proportional to $(f_{\mathrm{NL}})^4A_{\zeta}^4$ in Eq.~(\ref{eq:Ome}) grows, leading to the upper limit of $A_{\zeta}$ from Eq.~(\ref{eq:Ome}) becoming lower than that from Eq.~(\ref{eq:Otot}).

\section{Primordial black holes}\label{sec:3.0}
On small scales, large-amplitude primordial curvature perturbations can generate \acp{SIGW} upon re-entering the horizon after inflation. This process is inevitably accompanied by the formation of \acp{PBH}. In this section, we consider \acp{PBH} formed from large-amplitude primordial curvature perturbations on small scales, and the constraints they impose on small-scale primordial power spectrum and primordial non-Gaussianity. Furthermore, binary \ac{PBH} mergers will also generate an additional \ac{SGWB} \cite{Wang:2019kaf}. By combining the results of \acp{SIGW} from the previous section, we analyze the impact of binary \ac{PBH} mergers on the observations of \ac{SGWB}.

\subsection{Abundance of \acp{PBH}}\label{sec:3.1}
The abundance of \acp{PBH} can be expressed as \cite{Sasaki:2018dmp}
\begin{eqnarray}\label{eq:pbh1}
   f_{\mathrm{pbh}} \simeq 2.5 \times 10^8 \beta\left(\frac{g_*^{\text {form }}}{10.75}\right)^{-\frac{1}{4}}\left(\frac{m_{\mathrm{pbh}}}{M_{\odot}}\right)^{-\frac{1}{2}} \ .
\end{eqnarray}
 We consider an approximate formula for $f_{\mathrm{PBH}}$ in Eq.~(\ref{eq:pbh1}). The corresponding rigorous expressions are provided in Refs.~\cite{Ferrante:2022mui,Iovino:2024tyg,Franciolini:2023pbf}. The mass of \ac{PBH} is characterised by the scaling law relation 
\begin{equation}
\begin{aligned}
&M_{\mathrm{PBH}}=\mathcal{K} M_H\left(\mathcal{C}-\mathcal{C}_{\mathrm{th}}\right)^\gamma \ , \\
&M_H\approx 17 M_{\odot}\left(\frac{k}{10^6 \mathrm{Mpc}^{-1}}\right)^{-2}\left(\frac{g_*}{10.75}\right)^{-1 / 6} \ ,
\end{aligned}
\end{equation}
where the threshold $\mathcal{C}_{\text {th}}$ have been studied in Ref.~\cite{Musco:2020jjb}. The compaction function $\mathcal{C}=\mathcal{C}_1-\mathcal{C}_1^2 /(4 \Phi)$ can be obtained from the linear $\mathcal{C}_1=\mathcal{C}_{\mathrm{G}} \dif F / \dif \zeta_{\mathrm{G}}$ component, that uses $\mathcal{C}_{\mathrm{G}}=$ $-2 \Phi r \zeta_G^{\prime}$, where $\Phi=3(1+w)/(5+3w)$. In the case of local-type primordial non-Gaussianity, $F\left( \zeta_{\mathrm{G}} \right)=\zeta_{\mathrm{G}}+\frac{3}{5} f_{\mathrm{NL}} \zeta_{\mathrm{G}}^2$. We set $\mathcal{K}=4.4$ and $\gamma=0.38$ \cite{Musco:2023dak,Iovino:2024tyg}. The mass fraction $\beta$ can be obtained by integrating the probability distribution function
\begin{eqnarray}\label{eq:betat}
    \beta=\int_{\mathcal{D}} \mathcal{K}\left(\mathcal{C}-\mathcal{C}_{\mathrm{th}}\right)^\gamma \mathrm{P}_{\mathrm{G}}\left(\mathcal{C}_{\mathrm{G}}, \zeta_{\mathrm{G}}\right) \dif \mathcal{C}_{\mathrm{G}}  \dif \zeta_{\mathrm{G}} \ ,
\end{eqnarray}
where the domain of integration in Eq.~(\ref{eq:betat}) is $\mathcal{D}=$ $\left\{\mathcal{C}\left(\mathcal{C}_{\mathrm{G}}, \zeta_{\mathrm{G}}\right)>\mathcal{C}_{\text {th}} \wedge \mathcal{C}_1\left(\mathcal{C}_{\mathrm{G}}, \zeta_{\mathrm{G}}\right)<2 \Phi\right\}$. In Eq.~(\ref{eq:betat}), the Gaussian components are distributed as
\begin{equation}\label{eq:PG1}
    P_{\mathrm{G}}\left(\mathcal{C}_{\mathrm{G}}, \zeta_{\mathrm{G}}\right)=\frac{e^{\left[-\frac{1}{2\left(1-\gamma_{c r}^2\right)}\left(\frac{\mathcal{C}_{\mathrm{G}}}{\sigma_c}-\frac{\gamma_{c r} \zeta_{\mathrm{G}}}{\sigma_r}\right)^2-\frac{\zeta_\mathrm{G}^2}{2 \sigma_r^2}\right]}}{2 \pi \sigma_c \sigma_r \sqrt{1-\gamma_{c r}^2}} \ .
\end{equation}
The correlators in Eq.~(\ref{eq:PG1}) are given by
\begin{equation}\label{eq:sgm3}
\begin{aligned}
& \sigma_c^2=\frac{4 \Phi^2}{9} \int_0^{\infty} \frac{\dif k}{k}\left(k r_m\right)^4 W^2\left(k, r_m\right) P_\zeta^T \ , \\
& \sigma_{c r}^2=\frac{2 \Phi}{3} \int_0^{\infty} \frac{\dif k}{k}\left(k r_m\right)^2 W\left(k, r_m\right) W_s\left(k, r_m\right) P_\zeta^T \ , \\
& \sigma_r^2=\int_0^{\infty} \frac{\dif k}{k} W_s^2\left(k, r_m\right) P_\zeta^T \ ,
\end{aligned}
\end{equation}
where $P_\zeta^T=T^2\left(k, r_m\right) P_\zeta(k)$, and $\gamma_{c r} \equiv \sigma_{c r}^2 / \sigma_c \sigma_r$. Here, we have defined $W\left(k, r_m\right), W_s\left(k, r_m\right)$ and $T\left(k, r_m\right)$ as the top-hat window function, the spherical-shell window function, and the radiation transfer function \cite{Young:2022phe}. It is important to note that the calculation of \acp{PBH} abundance is highly model-dependent, and different models can lead to significant variations in the estimated abundance \cite{Iovino:2024tyg}.

\begin{figure}[htbp]

    \centering
    \includegraphics[width=\columnwidth]{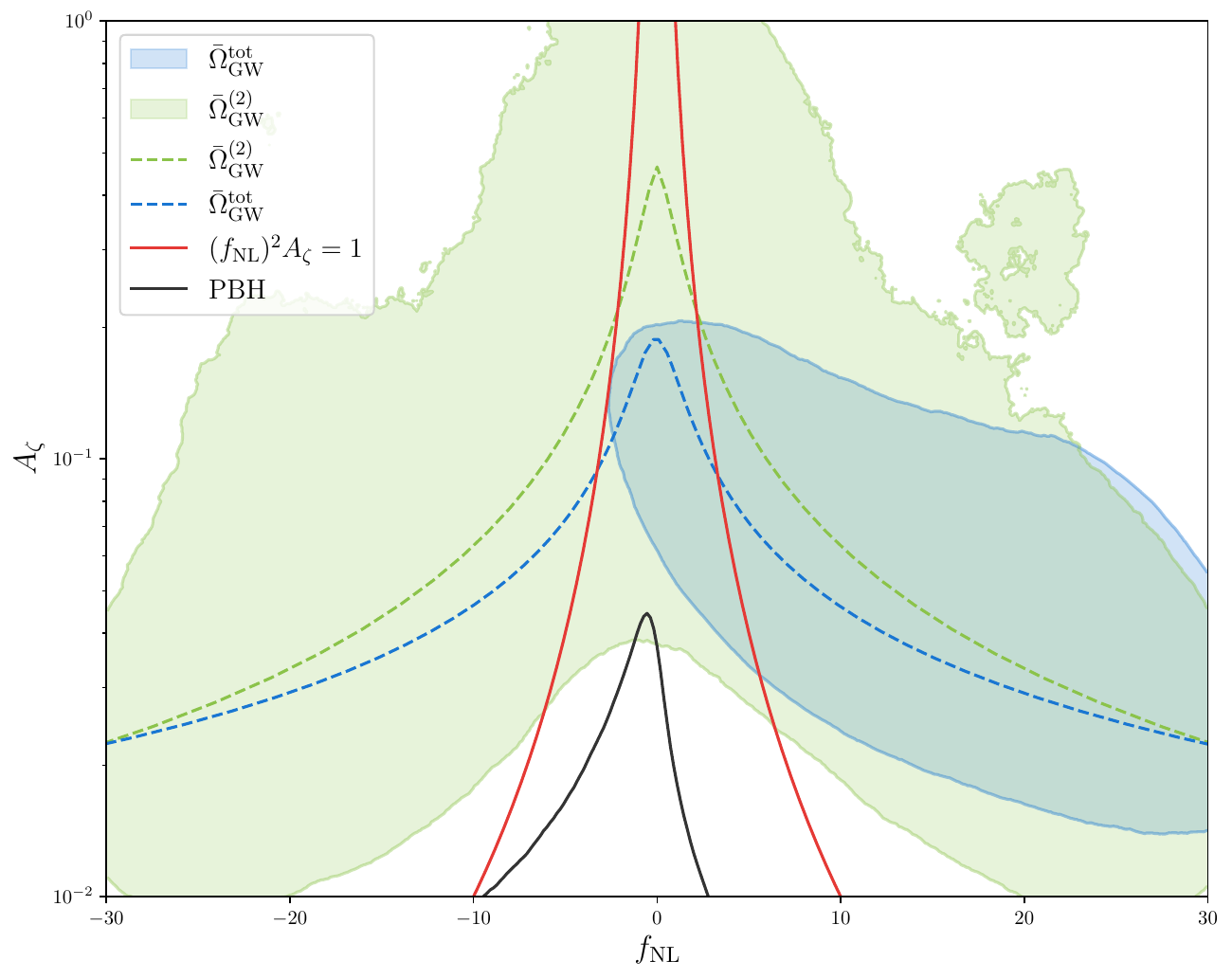}
    
\caption{\label{fig:constrain_LN} The constraints on $A_\zeta$ and $f_{\mathrm{NL}}$ assuming the \ac{LN} primordial power spectrum. The blue and green shaded regions represent the $68\%$ credible intervals corresponding to the two-dimensional posterior distribution shown in Fig.~\ref{fig:corner_eq3vseq10} with the \ac{KDE} method. The blue and green dashed lines indicate the upper bounds on the amplitude $A_\zeta$ shown in Fig.~\ref{fig:cosmo}. The red curve corresponds to the relation $(f_{\mathrm{NL}})^2 A_\zeta = 1$. The black line denotes the abundance of \acp{PBH} assuming $f_{\mathrm{PBH}}=1$.}
\end{figure}

As shown in Fig.~\ref{fig:constrain_LN}, we present the constraints on the amplitude of the small-scale primordial power spectrum $A_{\zeta}$ and the primordial non-Gaussian parameter $f_{\mathrm{NL}}$ in the case of the \ac{LN} primordial power spectrum, based on current \ac{CMB}$+$\ac{BAO}$+$\ac{PTA}$+$\acp{PBH} observations. Furthermore, in the case of local-type primordial non-Gaussianity, the second-order energy density $\bar{\Omega}^{(2)}_{\mathrm{GW}}$ is proportional to $A_{\zeta}^2+(f_{\mathrm{NL}})^2A_{\zeta}^3+(f_{\mathrm{NL}})^4A_{\zeta}^4$ , and each increase in the non-Gaussian order introduces an additional perturbative expansion coefficient $(f_{\mathrm{NL}})^2A_{\zeta}$.  This result holds for $n$-th order \acp{SIGW}, such that: $\bar{\Omega}^{(n)}_{\mathrm{GW}}\sim A_{\zeta}^n\sum_{m=0}^{n}(f_{\mathrm{NL}})^{2m}A_{\zeta}^m$. Therefore, the necessary condition for ensuring the convergence of the perturbative expansion of non-Gaussian contributions is $(f_{\mathrm{NL}})^2A_{\zeta}<1$. This theoretical constraint corresponds to the region below the red curve in Fig.~\ref{fig:constrain_LN}.

\begin{tcolorbox}
  \textbf{   In summary, the parameter space allowed by current observations corresponds to the region below the curves in Fig.~\ref{fig:constrain_LN}. When \acp{SIGW} dominate the current \ac{PTA} observations, the parameters $f_{\mathrm{NL}}$ and $A_{\zeta}$ must also reside within the blue or green shaded regions in Fig.~\ref{fig:constrain_LN}. }
\end{tcolorbox}

Through various types of cosmological observations across different scales, we can effectively constrain the primordial non-Gaussian parameter $f_{\mathrm{NL}}$ on small scales. It is important to note that these constraints depend on the specific form of the small-scale primordial power spectrum. In Sec.~\ref{sec:4.0}, we will analyze the impact of different primordial power spectra on the constraints of the parameter $f_{\mathrm{NL}}$. Additionally, the parameter space shown in Fig.~\ref{fig:constrain_LN} is influenced by experimental observations, and future high-precision cosmological measurements will further refine our ability to constrain small-scale primordial non-Gaussian parameter.

\subsection{\ac{SGWB} from binary \ac{PBH} mergers}\label{sec:3.2}
\begin{figure}[htbp]
\centering
\includegraphics[width=\linewidth]{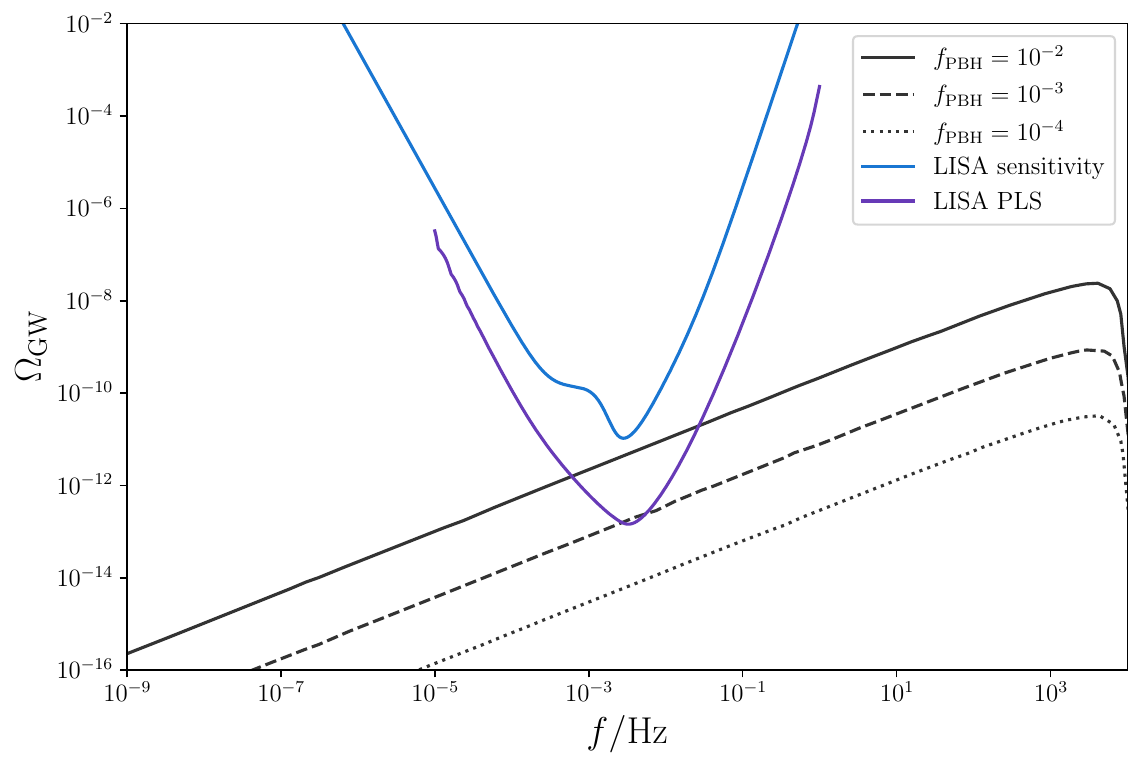}
\caption{The energy density spectrum $\bar{\Omega}^{\mathrm{PBH}}_{\mathrm{GW}}(k)$ in Eq.~(\ref{eq:PBHGW}). The black curves represent the energy spectra of ${\Omega}^{\mathrm{PBH}}_{\mathrm{GW}}(k)$ with $M_{\mathrm{PBH}}=1M_{\odot}$ and $f_{\mathrm{PBH}}=10^{-2}$ (solid), $10^{-3}$ (dashed) and $10^{-4}$ (dotted) \cite{Kohri:2020qqd}. The blue curve and the purple curve represent the stardand \ac{LISA} sensitivity and the \acf{PLS} \cite{Babak:2021mhe} with \ac{SNR} $=10$. \label{fig:pbh_spectrum}}
\end{figure}

In Sec.~\ref{sec:2.2}, we investigated how \acp{SIGW} influence the \ac{SNR} of \ac{LISA}. As we previously mentioned, when \acp{SIGW} dominate \ac{PTA} observations, their generation inevitably accompanies the formation of \acp{PBH}, and binary \ac{PBH} mergers also contribute to the \ac{SGWB}. In this paper, we follow the formation scenario of \ac{PBH} binaries proposed in Refs.~\cite{Wang:2016ana,Sasaki:2016jop,LIGOScientific:2018glc}. Based on the merger rate of the \ac{PBH} binaries, we can calculate the energy density spectra $\Omega_{\mathrm{GW}}^{\mathrm{PBH}}(f)$ of the corresponding \ac{SGWB}. For the \ac{SGWB} produced by  binary \ac{PBH} mergers, $\Omega_{\mathrm{GW}}^{\mathrm{PBH}}(f)$ can be expressed as 
\begin{equation}\label{eq:PBHGW}
\Omega_{\mathrm{GW}}^{\mathrm{PBH}}(f)=\frac{\nu}{\rho_{\mathrm{c}}} \int_0^{\frac{f_{\mathrm{cut}}}{f}-1} \frac{R_{\mathrm{PBH}}(z)}{(1+z) H(z)} \frac{\dif E_{\mathrm{GW}}}{\dif f_{\mathrm{s}}}\left(f_{\mathrm{s}}\right) \dif z \ ,
\end{equation}
where the explicit expressions of $\frac{\dif E_{\mathrm{GW}}}{\dif f_{\mathrm{s}}}\left(f_{\mathrm{s}}\right)$ and $R_{\mathrm{PBH}}(z)$ can be found in Ref.~\cite{Wang:2019kaf}. As shown in Fig.~\ref{fig:pbh_spectrum}, when \acp{SIGW} lie within the observational frequency range of \ac{PTA}, the \ac{SGWB} generated by the corresponding binary \ac{PBH} mergers falls within LISA's detection frequency band, which may affect the \ac{SNR} of \ac{LISA}.

\section{Shape of the primordial power spectra}\label{sec:4.0}
In the above discussion, we have provided the constraints on the local-type non-Gaussian parameter $f_{\mathrm{NL}}$ from current cosmological observations for a \ac{LN} primordial power spectrum. This result depends on the specific form of the primordial power spectrum. To study the effects of different shapes of the primordial power spectrum on small-scale cosmological observations, we need to combine current \ac{PTA} observational data and analyze the Bayesian factors of different primordial power spectra. Besides the previously analyzed \ac{LN} primordial power spectrum, we consider the \ac{BPL} primordial power spectrum \cite{You:2023rmn,Byrnes:2018txb}
\begin{equation}\label{eq:BPL}
\mathcal{P}_\zeta^{\mathrm{BPL}}(k)= \frac{A_{\zeta}(\alpha+\beta)}{\left(\beta\left(k / k_*\right)^{-\alpha }+\alpha\left(k / k_*\right)^{\beta }\right)}  \ ,
\end{equation}
and the $\delta$ (monochromatic) primordial power spectrum \cite{Cai:2018tuh}
\begin{equation}\label{eq:delta}
\mathcal{P}_\zeta^{\delta}(k)= A_{\zeta}k_*\delta\left( k-k_* \right)  \ .
\end{equation}

\begin{figure}[htbp]
    \captionsetup{
      justification=raggedright,
      singlelinecheck=true
    }
    \centering
	\subfloat[The constraints on $A_\zeta$ and $f_{\mathrm{NL}}$ assuming the \ac{BPL} primordial power spectrum..\label{fig:constrain_bpl}]  {\includegraphics[width=\columnwidth]{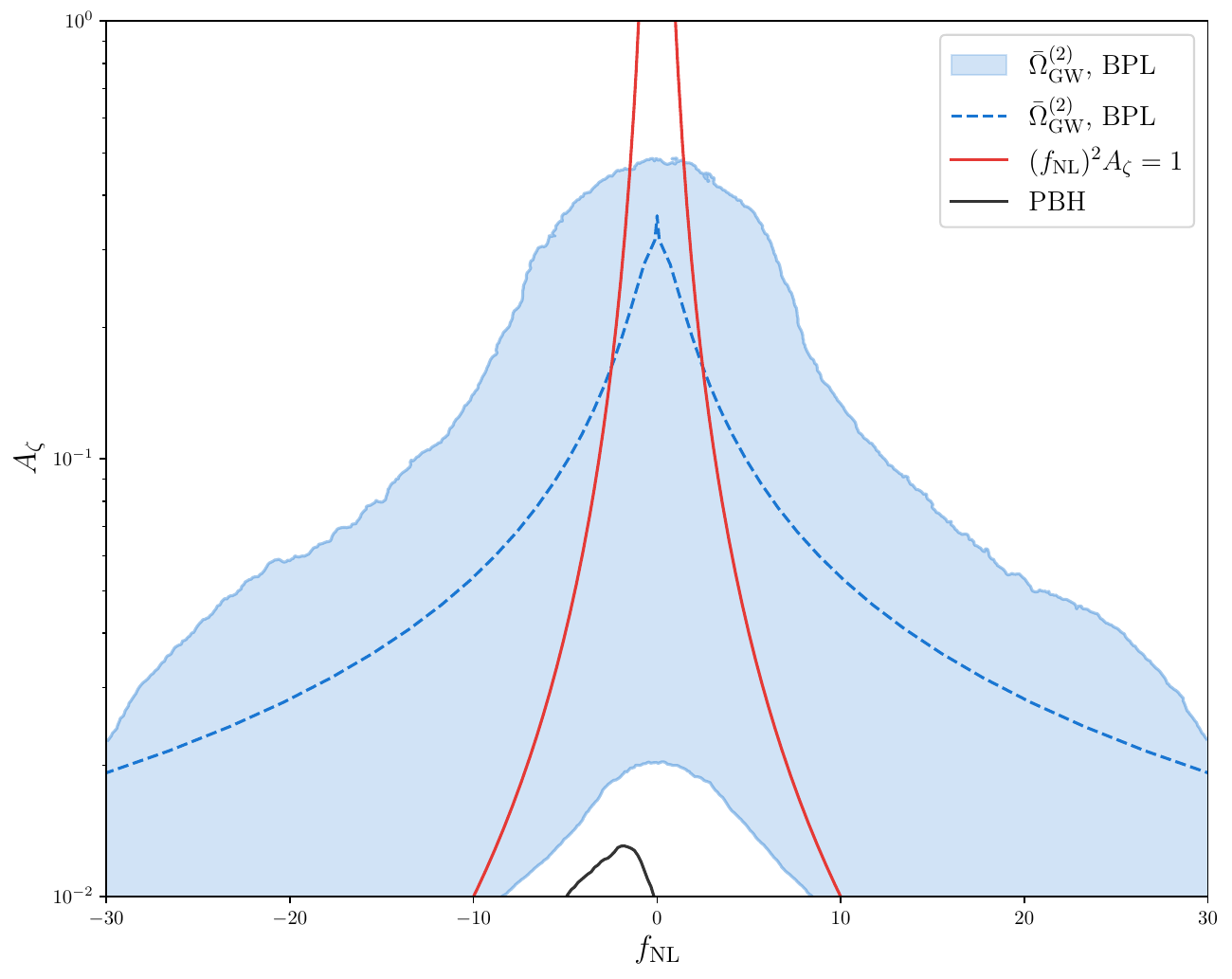}}
    \hspace{0.05cm}
        \subfloat[The constraints on $A_\zeta$ and $f_{\mathrm{NL}}$ assuming the monochromatic primordial power spectrum.\label{fig:constrain_mono}]{\includegraphics[width=\columnwidth]{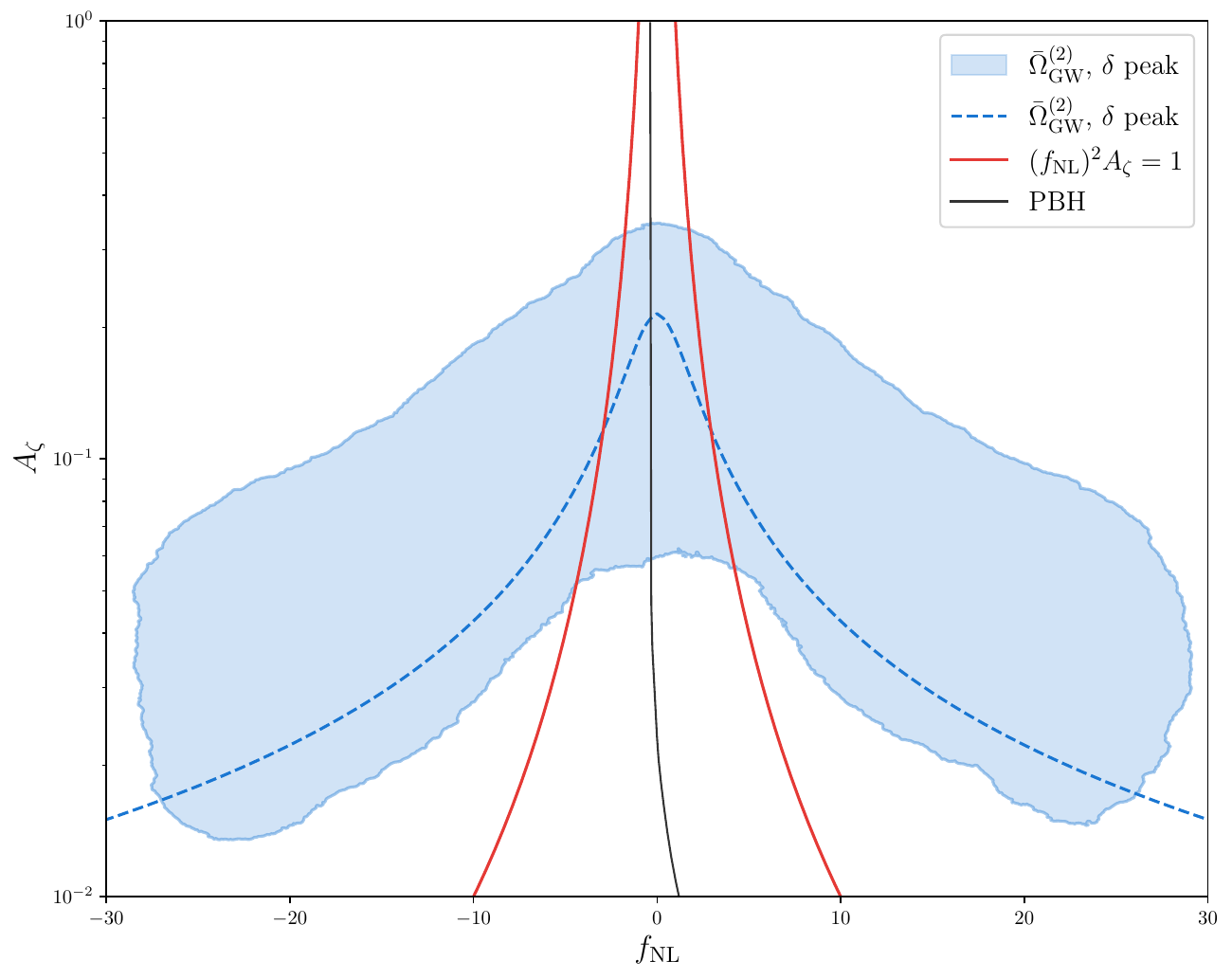}} 
\caption{\label{fig:constrain} The constraints on $A_\zeta$ and $f_{\mathrm{NL}}$ assuming different primordial power spectrum. The blue shaded regions represent the $68\%$ credible intervals corresponding to the two-dimensional posterior distribution with \ac{KDE} method. The blue dashed lines indicate the upper bounds on the amplitude $A_\zeta$ from \ac{CMB}$+$\ac{BAO} data. The red curve corresponds to the relation $(f_{\mathrm{NL}})^2 A_\zeta = 1$. The black line denotes the abundance of \acp{PBH} assuming $f_{\mathrm{PBH}}=1$. As stated in Ref.~\cite{Ferrante:2022mui}, Eq.~(\ref{eq:betat}) is not applicable to a monochromatic primordial power spectrum. Therefore, we adopt the approach proposed in Ref.~\cite{Byrnes:2012yx} to calculate the \ac{PBH} abundance under the monochromatic primordial power spectrum scenario.}
\end{figure}

Similar to the discussions in Sec.~\ref{sec:2.0} and Sec.~\ref{sec:3.0}, the constraints imposed by current cosmological observations on the parameter space of the \ac{BPL} primordial power spectrum and the monochromatic primordial power spectrum are presented in Fig.~\ref{fig:constrain_bpl} and Fig.~\ref{fig:constrain_mono}, respectively. The corresponding posterior distributions are given in the Appendix.~\ref{sec:A}. For the monochromatic power spectrum, the prior distributions of $\log(f_*/\mathrm{Hz})$, $\log(A_\zeta)$, and $f_{\mathrm{NL}}$ are uniformly distributed over the ranges $[-10,-5]$, $[-3,0]$, and $[-30,30]$ respectively. For the \ac{BPL} power spectrum, the prior distributions of $\log(f_*/\mathrm{Hz})$, $\log(A_\zeta)$, $\alpha$, $\beta$ and $f_{\mathrm{NL}}$ are uniformly distributed over the ranges $[-10,-5]$, $[-3,0]$, $[1,3]$, $[1,3]$,and $[-30,30]$ respectively. 

\begin{figure*}[htbp]
    \centering
    \includegraphics[width=1.8\columnwidth]{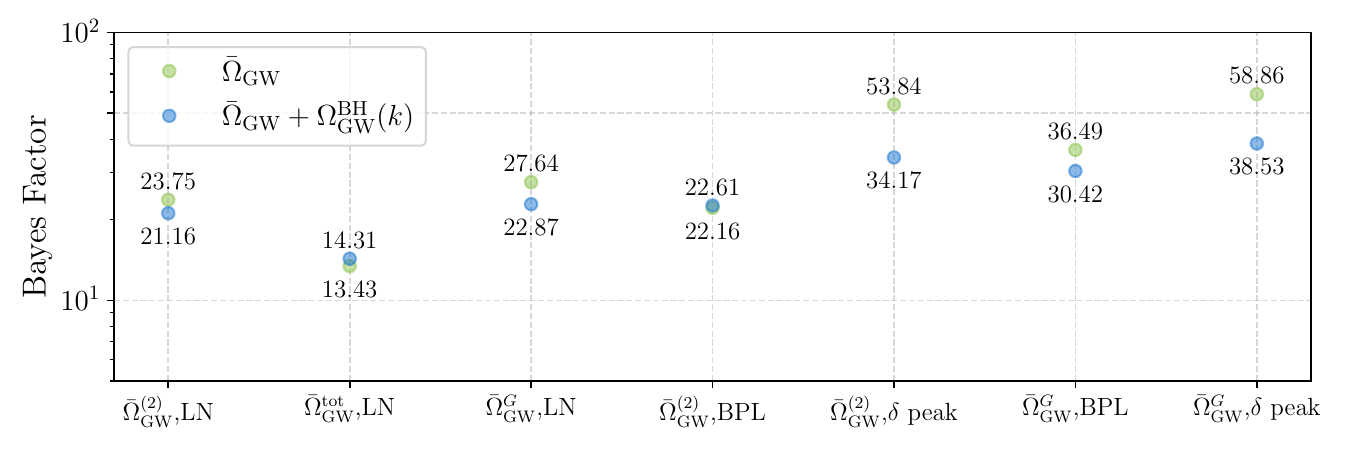}
\caption{\label{fig:bayes} The Bayes factors between different models. The vertical axis represents the Bayes factor of different models relative to \ac{SMBHB}, and the horizontal axis represents the different models. The green dots are for models without \ac{SMBHB} and the blue dots are for models in combination with the \ac{SMBHB} signal. }
\end{figure*}

To rigorously quantify the probability of different models dominating current \ac{PTA} observations, we employ the Bayes factor to compare them. Specifically, the Bayes factor is defined as $B_{i,j} = \frac{Z_i}{Z_j}$, where $Z_i$ represents the evidence of model $H_i$. Fig.~\ref{fig:bayes} illustrates Bayes factors for comparisons between various models and the \ac{SMBHB} model. As shown in Fig.~\ref{fig:bayes}, the Bayesian factor analysis indicates that \acp{SIGW} generated by the monochromatic primordial power spectrum are more likely to dominate the current \ac{PTA} observations. 

In Table.~\ref{ta:1}, we summarize the main results of this section. The observational constraints on the parameter of small-scale primordial non-Gaussianity $f_{\mathrm{NL}}$ depend not only on the specific form of the primordial power spectrum but also on its amplitude $A_{\zeta}$. In Table.~\ref{ta:1}, we have set $A_{\zeta}=10^{-2}$. Note that the above conclusion differs from the constraints on $f_{\mathrm{NL}}$ at large scales, as the primordial power spectrum at large scales has already been strictly constrained by large-scale cosmological observations. Therefore, different forms of the primordial power spectrum do not need to be considered when determining the constraints on $f_{\mathrm{NL}}$ at large scales. 

Furthermore, if no specific assumption is made about the origin of the \ac{SGWB} in the \ac{PTA} frequency band, the energy density spectrum of second-order \acp{SIGW} must not exceed the observed energy density spectrum in the \ac{PTA} band. In this scenario, current \ac{PTA} observations can still serve as an upper bound on the energy density spectrum of second-order \acp{SIGW}. Thus, when we require \acp{SIGW} to dominate the current \ac{PTA} observations, the parameter space of $f_{\mathrm{NL}}$ and $A_{\zeta}$ must not only satisfy other cosmological constraints but also lie within the blue-shaded regions in Fig.~\ref{fig:constrain_bpl} and Fig.~\ref{fig:constrain_mono}. Since $\bar{\Omega}^{(2)}_{\mathrm{GW}}(k)$ increases with $f_{\mathrm{NL}}$ and $A_{\zeta}$, if we do not assume that \acp{SIGW} dominate the \ac{SGWB} in the \ac{PTA} frequency band, then the value of $A_{\zeta}$ for a given $f_{\mathrm{NL}}$ must not exceed the blue-shaded regions in Fig.~\ref{fig:constrain_bpl} and Fig.~\ref{fig:constrain_mono}.

\begin{table}[h!]
\centering
\begin{tabular}{lcc}
\hline \hline & $f_{\mathrm{NL}}$ & Bayesian factors \\
\hline \ac{LN}  & $-9.5<f_{\mathrm{NL}}<2.9$  & 23.75 \\
\rule{0pt}{12pt}
\ac{BPL} & $-5.0<f_{\mathrm{NL}}<-0.1$  & 22.61 \\
\rule{0pt}{12pt}
$\delta$ peak & $-10.0<f_{\mathrm{NL}}<1.2$  & 53.84 \\
\rule{0pt}{12pt}
\ac{SMBHB} & $\times$  & 1 \\
\hline \hline
\end{tabular}
\caption{Constraints from current cosmological observations on small-scale local-type non-Gaussian parameters. We set $A_{\zeta}=10^{-2}$.}
\label{ta:1}
\end{table}

Moreover, as discussed in Sec.~\ref{sec:2.0}, the cross-correlation function $\langle h^{\lambda,(3)}_{\mathbf{k}} h^{\lambda',(2)}_{\mathbf{k}'} \rangle\sim f_{\mathrm{NL}}A_{\zeta}^3$ can impact the results in Table.~\ref{ta:1}. When constraining the parameter $f_{\mathrm{NL}}$ using the energy density spectrum in Eq.~(\ref{eq:Otot}), the current \ac{PTA} observations cannot be dominated by \acp{SIGW}. As indicated by the blue shaded region in Fig.~\ref{fig:constrain_LN}, the presence of the cross-correlation function $\langle h^{\lambda,(3)}_{\mathbf{k}} h^{\lambda',(2)}_{\mathbf{k}'} \rangle$ suppresses the total energy density spectrum of \acp{SIGW} when $f_{\mathrm{NL}}$ is negative, thus excluding certain parameter regions based on cosmological constraints. In this scenario, insisting that \acp{SIGW} dominate the current \ac{PTA} observations would contradict the upper bound on the primordial black hole abundance.

\section{Conclusion and discussion}\label{sec:5.0}
To study the constraints imposed by current cosmological observations on the small-scale primordial power spectrum $\mathcal{P}_{\zeta}\left(k \right)$ and the local-type non-Gaussian parameter $f_{\mathrm{NL}}$, we analyzed the impact of large-amplitude small-scale primordial curvature perturbations, which induce the formation of \acp{PBH} and corresponding second-order \acp{SIGW}, on current cosmological observations at different scales. Future, more precise observations from \ac{PTA}, \acp{PBH}, and large-scale cosmological observations will enable us to more tightly constrain the parameter space of $\mathcal{P}_{\zeta}\left(k \right)$ and $f_{\mathrm{NL}}$. Furthermore, since \acp{SIGW} and \acp{PBH} abundance depend on the shape of the primordial power spectrum, we studied the impact of different primordial power spectra on the constraints of the parameter space and rigorously analyzed the Bayesian factors of different models. Moreover, the constraints on $f_{\mathrm{NL}}$ imposed by current cosmological observations for different forms of the primordial power spectrum are summarized in Table~\ref{ta:1}. The constraints on $f_{\mathrm{NL}}$ presented in Table.~\ref{ta:1} depend on the specific form and the amplitude of the primordial power spectra. 

In this study, we evaluated the influence of second-order and third-order \acp{SIGW} on current cosmological observations. As shown in Refs.~\cite{DeLuca:2023tun,Nakama:2015nea,Chang:2022aqk,Nakama:2016enz,Zhou:2023itl}, second-order density perturbations induced by primordial perturbations will have a significant impact on the threshold $\delta_c$ and probability distribution function $P\left(\delta \right)$ of \acp{PBH}. When studying the large-amplitude primordial perturbations on small scales, the effects of higher-order cosmological perturbations cannot be ignored \cite{Zhou:2024ncc}.  In addition to the influence of higher-order corrections, several other effects could impact the parameter space of $\mathcal{P}_{\zeta}\left(k \right)$ and $f_{\mathrm{NL}}$.  These include the effects of different dominant era of universe \cite{Chen:2024fir,Balaji:2023ehk,Zhu:2023gmx}, the impact of large-amplitude small-scale primordial tensor perturbations \cite{Chang:2022vlv,Bari:2023rcw,Yu:2023lmo,Picard:2024ekd,Picard:2023sbz}, and interactions between \acp{SIGW} and matter \cite{Saga:2014jca,Zhang:2022dgx,Yu:2024xmz,Sui:2024nip}. The impact of these physical effects on the constraints of primordial non-Gaussianity might be systematically investigated in future research.

{As discussed in Sec.~\ref{sec:4.0}, due to the uncertainty in the shape of the primordial power spectrum on small scales, the constraints from current cosmological observations on small-scale primordial non-Gaussianity are highly sensitive to the assumed form of the primordial power spectrum. In this paper, we consider three different forms of the primordial power spectrum: the \ac{LN} spectrum, the \ac{BPL} spectrum, and the monochromatic spectrum. Each of these can be realized through specific inflationary models. Specifically, Refs.~\cite{Zhou:2020kkf,Addazi:2022ukh,Peng:2021zon,Chen:2024gqn} show that the small-scale primordial spectra generated via resonant amplification can be approximately described by monochromatic or \ac{LN} forms, while Ref.~\cite{Atal:2021jyo} proposes an inflationary model that gives rise to the \ac{BPL} spectrum. However, for models based on modified gravity theories \cite{Pi:2017gih}, ultra-slow-roll scenarios \cite{Byrnes:2018txb}, or axion inflation \cite{Garcia-Bellido:2016dkw}, the resulting primordial power spectra generally do not conform to the three forms considered here. Given a specific inflationary model, the observational constraints discussed in this paper allow us to constrain its parameter space. Further research along these lines may be pursued in future work.}

\section*{Data availability}\label{sec:6.0}
The data that support the findings of this article are openly available in Ref. \cite{Nanograv:KDE}.

\vspace{0.3cm}
\begin{acknowledgements} 
This work has been funded by the National Nature Science Foundation of China under grant No. 12447127.  
\end{acknowledgements}

\appendix
\section{Posterior distributions}\label{sec:A}

Based on the theoretical results of the energy density spectrum of second-order\acp{SIGW}, we present the posterior distributions determined by current \ac{PTA} observations under different forms of the primordial power spectrum. As shown in Fig.~\ref{fig:bayes}, we further analyze the impact of \acp{SIGW} on \ac{PTA} observations when $f_{\mathrm{NL}}$. The posterior distributions for different models are provided in Fig.~\ref{fig:corner_eq10} $\sim$ Fig.~\ref{fig:corner_eq3_gaussian}. The contours in the off-diagonal panels denote the $68\% $ and $95 \%$ credible intervals of the 2D posteriors. The numbers above the figures represent the median values and $1$-$\sigma$ ranges of the parameters.

% \clearpage
\vspace{0.5cm}

\noindent
\begin{minipage}[t]{\linewidth}
    \centering
    \includegraphics[width=0.8\linewidth]{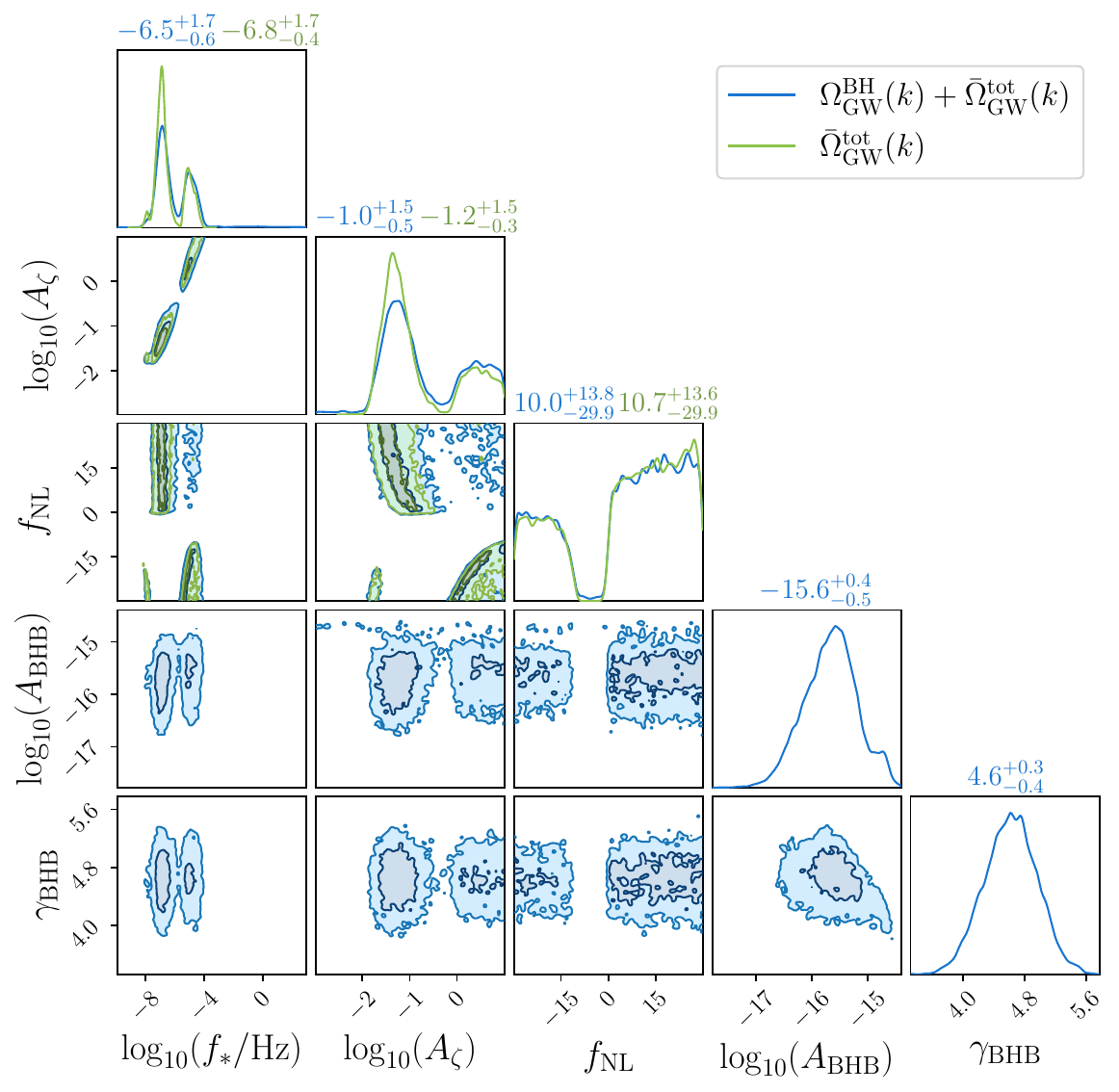}
    \captionof{figure}{\label{fig:corner_eq10} Posterior distributions of $\bar{\Omega}^{\mathrm{tot}}_{\mathrm{GW}}(k)$ and $\bar{\Omega}^{\mathrm{BH}}_{\mathrm{GW}}(k)+\bar{\Omega}^{\mathrm{tot}}_{\mathrm{GW}}(k)$ with \ac{LN} primordial power spectrum.}
\end{minipage}

\vspace{0.6cm}

\noindent
\begin{minipage}[t]{0.95\linewidth}
    \centering
    \includegraphics[width=\linewidth]{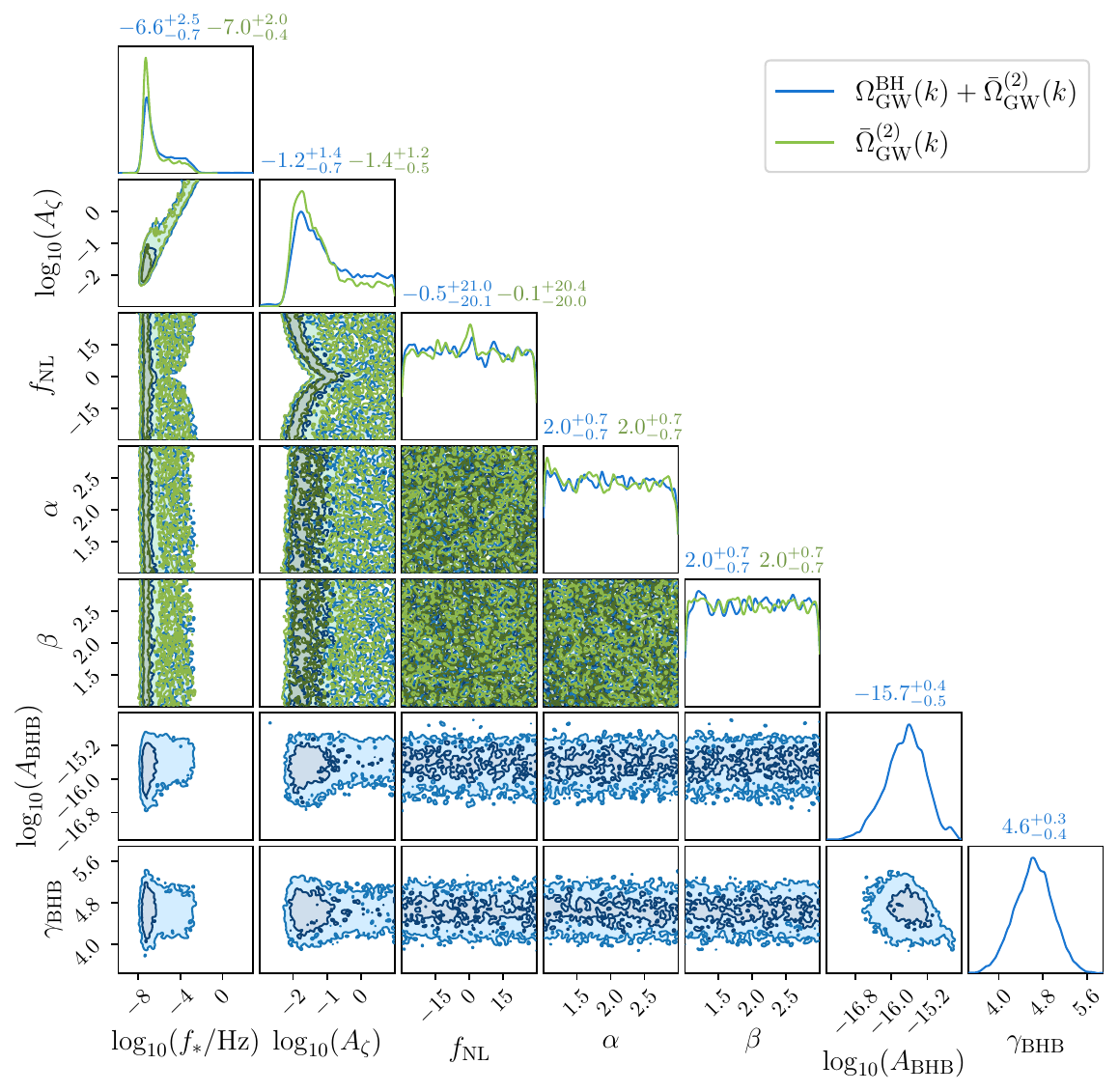}
    \captionof{figure}{\label{fig:corner_eq3_bpl} Posterior distributions of $\bar{\Omega}^{(2)}_{\mathrm{GW}}(k)$ and $\bar{\Omega}^{\mathrm{BH}}_{\mathrm{GW}}(k)+\bar{\Omega}^{(2)}_{\mathrm{GW}}(k)$ with \ac{BPL} primordial power spectrum.}
\end{minipage}%
\vspace{0.6cm}
\hfill
\begin{minipage}[t]{0.95\linewidth}
    \centering
    \includegraphics[width=\linewidth]{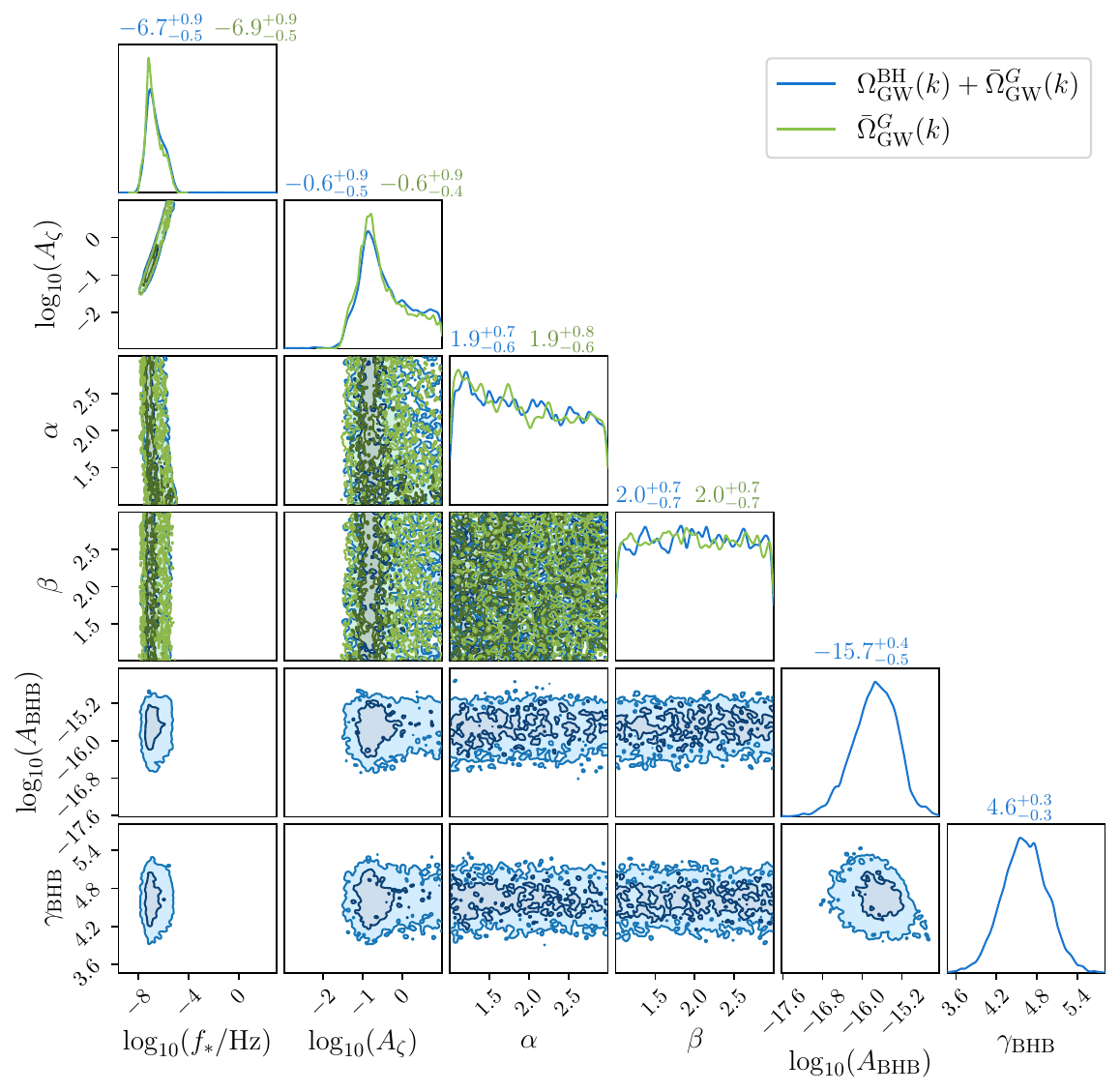}
    \captionof{figure}{\label{fig:corner_eq3_bpl_gaussian} Posterior distributions of $\bar{\Omega}^{G}_{\mathrm{GW}}(k)$ and $\bar{\Omega}^{\mathrm{BH}}_{\mathrm{GW}}(k)+\bar{\Omega}^{G}_{\mathrm{GW}}(k)$ with \ac{BPL} primordial power spectrum.}
\end{minipage}

\vspace{0.6cm}

\noindent
\begin{minipage}[t]{0.95\linewidth}
    \centering
    \includegraphics[width=\linewidth]{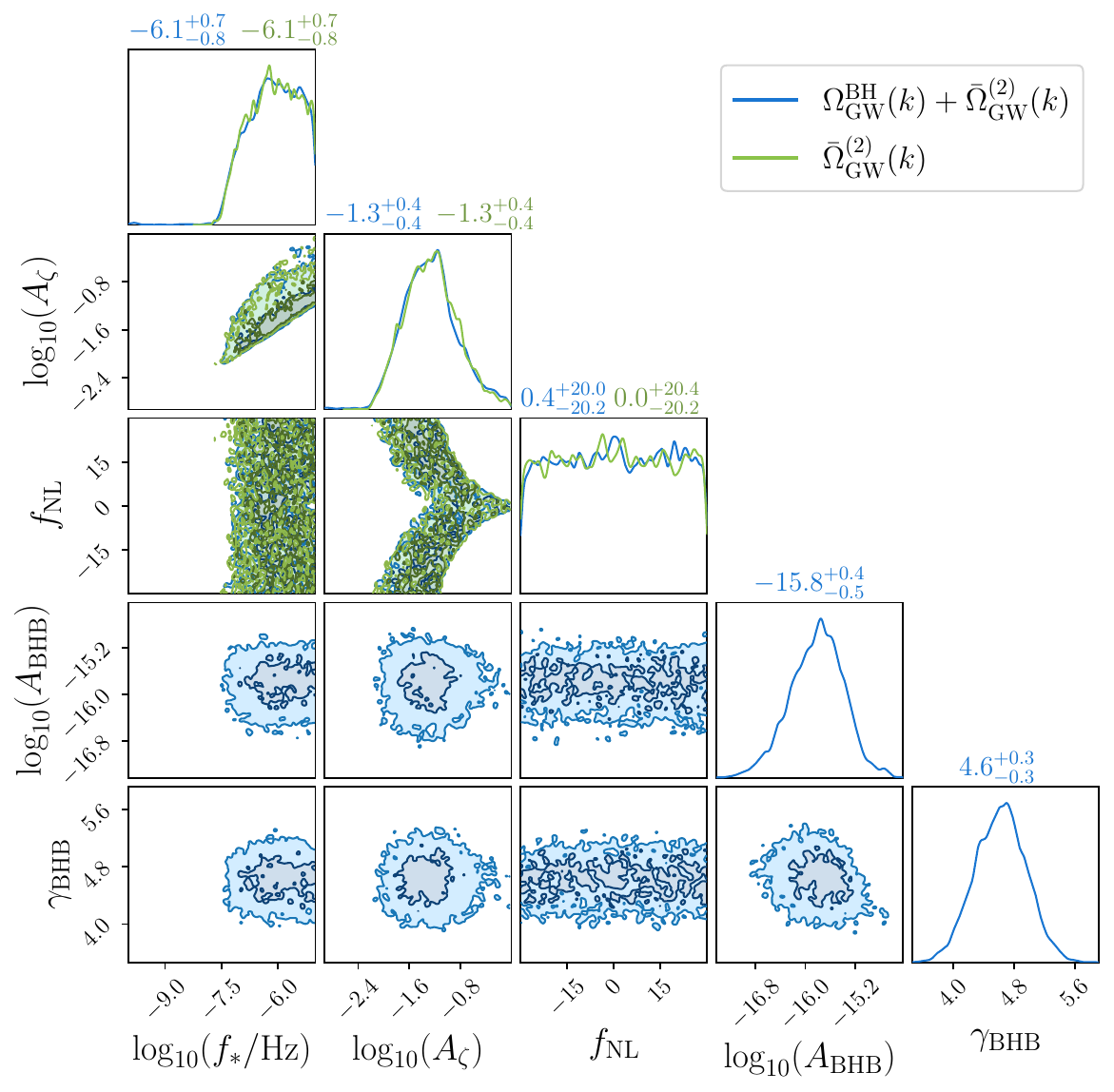}
    \captionof{figure}{\label{fig:corner_eq3_mono} Posterior distributions of $\bar{\Omega}^{(2)}_{\mathrm{GW}}(k)$ and $\bar{\Omega}^{\mathrm{BH}}_{\mathrm{GW}}(k)+\bar{\Omega}^{(2)}_{\mathrm{GW}}(k)$ with monochromatic primordial power spectrum.}
\end{minipage}%
\vspace{0.6cm}
\hfill
\begin{minipage}[t]{0.95\linewidth}
    \centering
    \includegraphics[width=\linewidth]{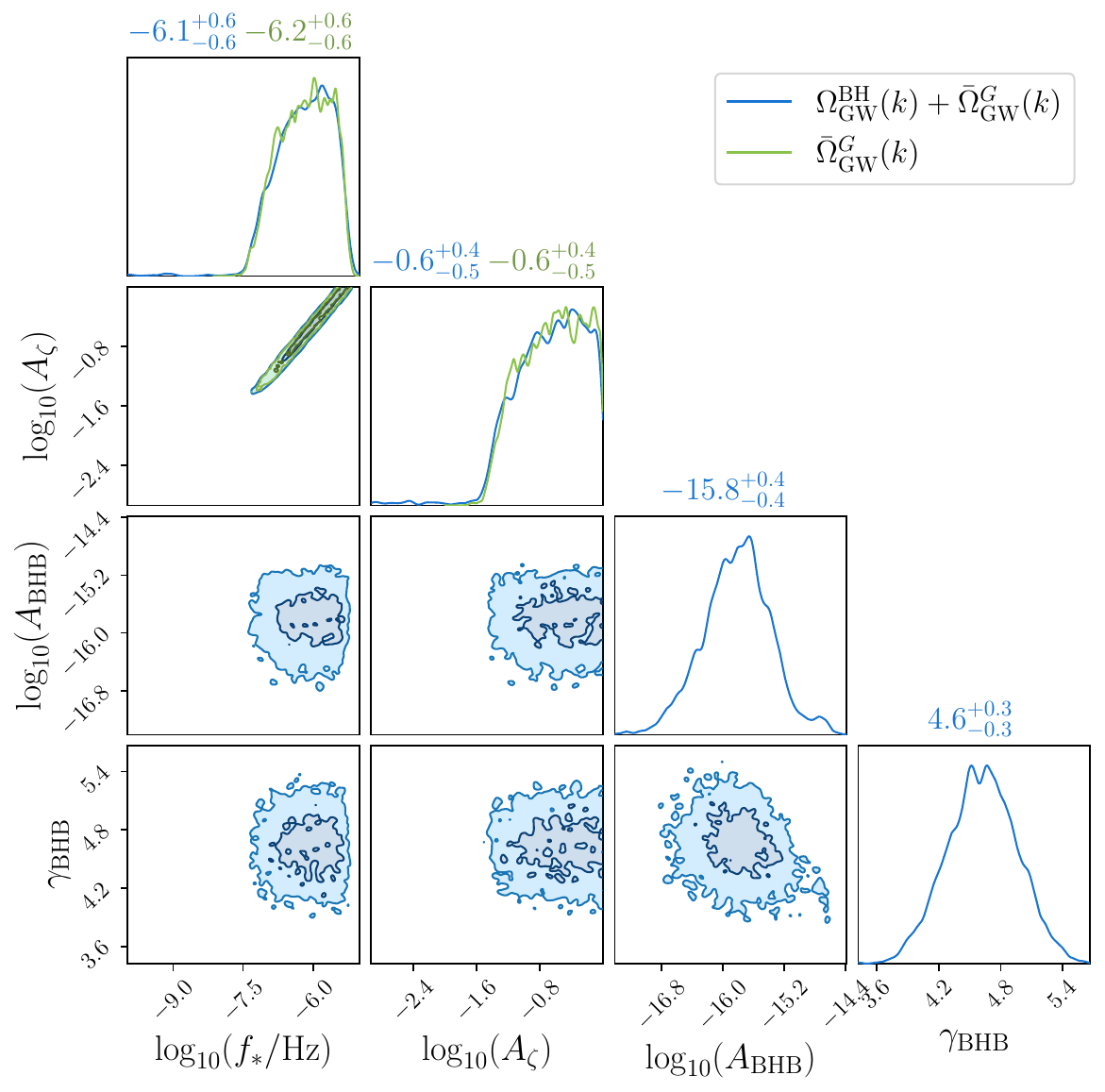}
    \captionof{figure}{\label{fig:corner_eq3_mono_gaussian} Posterior distributions of $\bar{\Omega}^{G}_{\mathrm{GW}}(k)$ and $\bar{\Omega}^{\mathrm{BH}}_{\mathrm{GW}}(k)+\bar{\Omega}^{G}_{\mathrm{GW}}(k)$ with monochromatic primordial power spectrum.}
\end{minipage}

\vspace{0.6cm}

%It is difficult to put this figure in the left column...
\clearpage  
\noindent
\begin{minipage}[H]{\linewidth}
    \centering
    \includegraphics[width=0.95\linewidth]{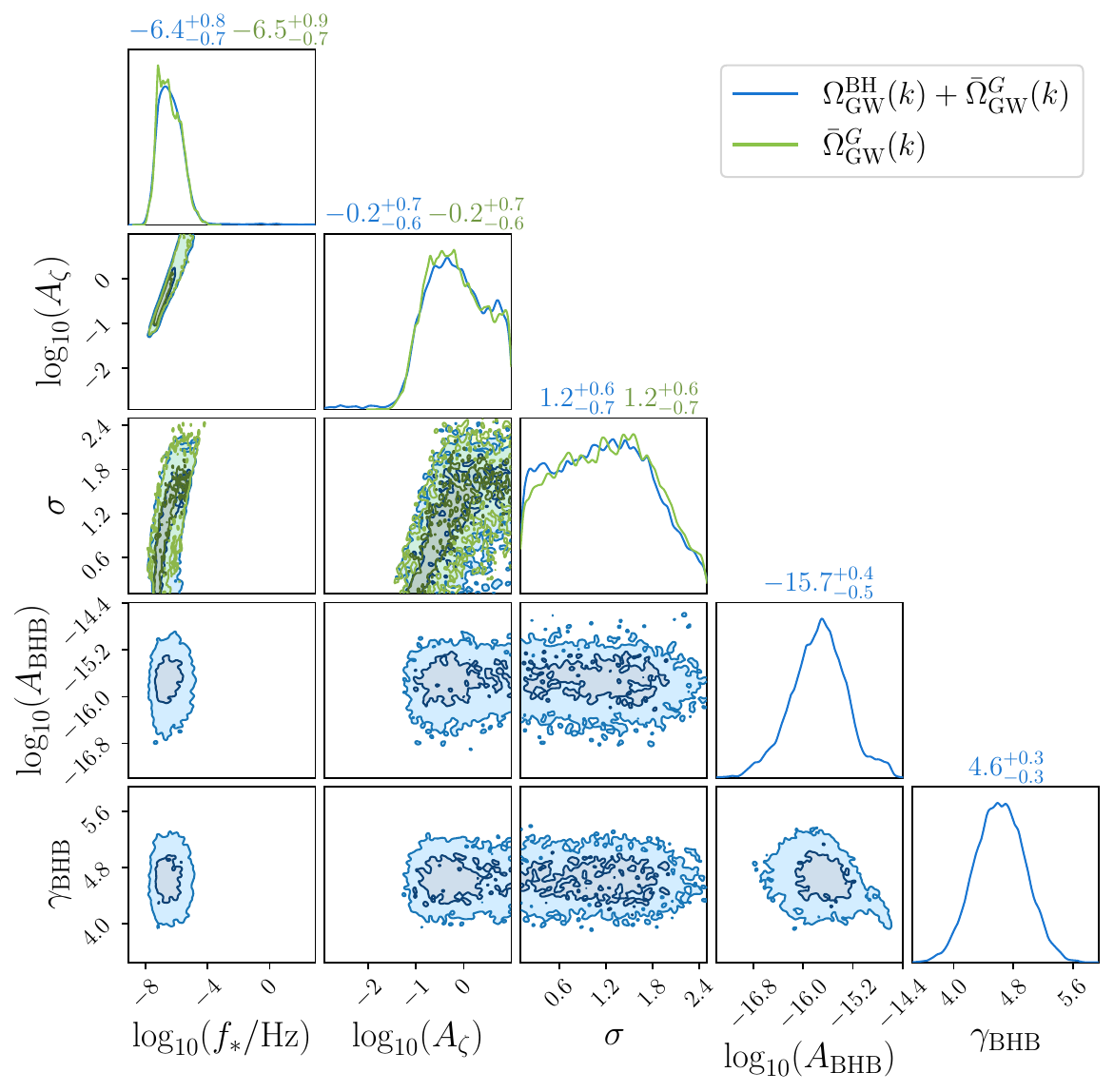}
    \captionof{figure}{\label{fig:corner_eq3_gaussian} Posterior distributions of $\bar{\Omega}^{G}_{\mathrm{GW}}(k)$ and $\bar{\Omega}^{\mathrm{BH}}_{\mathrm{GW}}(k)+\bar{\Omega}^{G}_{\mathrm{GW}}(k)$ with \ac{LN} primordial power spectrum.}
\end{minipage}

\vspace{1em}

\FloatBarrier

\bibliography{outputNotes}

\end{document}